\documentclass[prb,twocolumn,showpacs,amsfonts,amssymb,floats,superscriptaddress,aps]{revtex4}
\usepackage{graphicx}

\usepackage{wasysym,color}

\def\ket#1{|#1\rangle }
\def\bra#1{\langle#1 | }

\def\punkt{\;\; .}

\def\expect#1{\langle#1 \rangle}

\def\w{\omega}
\def\H{{\cal H}}
\def\e{\epsilon}
\def\s{\sigma}

\def\Tr#1{\textrm{Tr}\left[#1\right]}
\def\non{\nonumber\\ }

\def\feff{f_{\mbox{\small eff}}}

\newcommand{\ew}[1]{\langle #1\rangle}
\newcommand{\comment}[1]{}

\begin{document}

\title{Comparison between scattering-states numerical renormalization group and the 
  Kadanoff-Baym-Keldysh approach to quantum transport: Crossover from weak to strong 
  correlations}


\author{Sebastian Schmitt}
\author{Frithjof B. Anders}
\affiliation{Lehrstuhl f\"ur Theoretische Physik II, Technische Universit\"at Dortmund,
Otto-Hahn-Str. 4, 44221 Dortmund,
Germany}

\date{\today}

\begin{abstract}
The quantum transport through nanoscale junctions is governed by the
charging energy  $U$ of the device. 
We employ the recently developed scattering-states numerical renormalization group approach
to open quantum systems to study nonequilibrium Green's functions and current-voltage
characteristics of such junctions for small and intermediate values of $U$. 
We establish the accuracy of the approach by a comparison with diagrammatic
Kadanoff-Baym-Keldysh results which become exact in the weak coupling limit $U\to 0$.   
We demonstrate the limits of the diagrammatic expansions 
at intermediate values of the charging energy.  
While the numerical renormalization group approach correctly 
predicts only one single, universal low-energy  scale at zero bias voltage, 
some diagrammatic expansions yield  two  different  low-energy scales
for 
the magnetic and the charge fluctuations.
At large voltages, however, the self-consistent second Born as well as the GW approximation
reproduce the scattering-states renormalization group  spectral functions 
for symmetric junctions, while  for asymmetric junctions 
the voltage-dependent redistribution of spectral weight  differs significantly 
in the different approaches. The second-order perturbation theory does not 
capture the correct single-particle 
dynamics at large bias and violates current conservation for 
asymmetric junctions.
\end{abstract}
\maketitle

\section{Introduction}
\label{sec:1}

Quantum dots and single-molecule junctions have been considered as possible building blocks 
for nano-electronics and
quantum-information processing.\cite{Kouwenhoven2004,AbrahamNitzan2003,tautzKondoSTM08} 
Recent technological progress has made it possible to manufacture and study 
electron transport trough ultra-small quantum-dot devices, nanotubes or single 
molecules.\cite{NatureGoldhaberGordon1998,goldhaberSET98,wielNonEqKondo00,nagaokaTempDepKondo02,
  quayMagFieldQdot07,tautzKondoSTM08,grobisScalingQdot08,amashaKondoSplit05,liuMagFieldQdot09} 
These devices are designed as a small central region comprising of a quantum dot or a single molecule
which is coupled to at least two leads where the finite bias voltage is applied to. 
The investigation of  such devices is of fundamental importance for our understanding 
of open quantum systems out of equilibrium. 

Due to the quantization of the charge, the physical properties of such junctions
are dominated by many-body 
effects at temperatures below  the
charging energy $U=e^2/(2C)$, where $C$ is the 
capacitance of the device. 
The experimental devices are often fully controllable by external gate electrodes or 
elongation of the scanning tunneling microscope tip.\cite{tautzKondoSTM08} 
This gives the opportunity to directly study true many-body correlation effects, such as the 
Kondo effect (see for example Ref.\ \onlinecite{hewson:KondoProblem93} and \onlinecite{kondo40yearsSeries05}),
under the influence  of an external bias voltage.
However, the theoretical understanding of the interplay 
between coherent transport favored by many-body correlations
and current-driven dephasing at finite bias 
is  still at its infancy and further investigations are needed.

The present work has two main objectives. On the one hand, we establish the
reliability of the recently introduced scattering-states numerical renormalization group
(SNRG) approach\cite{AndersSSnrg2008}  to quantum transport
by comparing results for small values of $U$ to the diagrammatic Kadanoff-Baym-Keldysh  
approach,  which becomes exact in the limit $U\to 0$.
On the other hand, we will discuss discrepancies and  reveal shortcomings of those diagrammatic 
approaches at intermediate values of the charging energy.

We investigate quantum transport through a quantum-dot device using
a minimal model\cite{MeirWingreen1994}  where the  complex interacting region is replaced by 
a single spinful orbital which is coupled to two noninteracting leads.
A single  Coulomb matrix element $U$ accounts for the charging energy of the device.
We calculate nonequilibrium spectral functions\cite{AndersNeqGf2008}  
and current-voltage (IV) characteristics  using the SNRG
as well as  different approximations\cite{thygesenNonEqGW07,darancetNeqGW07,spataruGwMillis2009,GWvanLeeuwen2009}  
within the diagrammatic Kadanoff-Baym-Keldysh 
expansion in the local Coulomb interaction $U$.

Over the past 40 years, the Keldysh technique~\cite{Keldysh65} has proven to be 
the most successful approach to nonequilibrium dynamics. 
In the context of quantum transport through nano-junctions 
direct expansions in the
interaction\cite{yeyatiNonEqMPT93,takagiMagFieldSIAM99,matsumotoNeqSIAM00,fujiiPerturbKondoNeq03} 
as well as self-consistent
re-summation schemes have been employed.\cite{thygesenNonEqGW07,darancetNeqGW07,spataruGwMillis2009,GWvanLeeuwen2009}
However, such diagrammatic expansions rely on a small 
expansion parameter, and are, therefore, confined to
weak coupling. But quantum-impurity models\cite{BullaCostiPruschke2007}
commonly used in 
the theory of quantum transport on the molecular level often exhibit 
infra-red divergences in perturbation theory\cite{hewson:KondoProblem93} 
which also restrict the diagrammatic Keldysh approaches
to certain parameter regimes usually to high temperature or to large bias.

In contrast to equilibrium conditions, where complete
and accurate solutions can be obtained using a variety
of nonperturbative techniques such as the Bethe
{\em ansatz},\cite{BA-Kondo,BA-Anderson}  conformal field
theory,\cite{CFTa,AffleckLud93} or Wilson's numerical renormalization
group (NRG) approach,\cite{Wilson75,BullaCostiPruschke2007}
techniques  for calculating quantum-transport out
of equilibrium remain largely at the development stage. Recent
advancements on the analytical side
\cite{kaminskiQdotNeq00,roschNeqKondo01,jakobsNeqFRG07,grezziFRGNeq07,DoyonAndrei2005,MethaAndrei2005}
include suitable adaptations of the Wegner's 
flow-equation\cite{Kehrein2005,KehreinFlow2006} and the 
real-time renormalization-group
method.\cite{SchoellerKoenig2000,Schoeller2009a,PSS09}  
These methods can successfully access large voltages,  
but are generally 
confined to the weak-coupling regime. Based on the scattering-states 
approach to quantum transport\cite{Hershfield1993,DoyonAndrei2005}  
the Bethe ansatz was extended to  quantum-impurity models out of equilibrium,\cite{MethaAndrei2005} 
but remains limited to a certain class of models.

On the numerical side, progress has been made in several
directions. Currents have been extracted from time-dependent 
density matrix renormalization 
group\cite{Schmitteckert2004,SchollwoeckDMRG2005,kirinotDMRGSIAM08,meisnerRealTimeSIAM09}
calculations using finite 1D wires, and the results agree well with 
Bethe ansatz results for certain models.\cite{BoulatSaleurSchmitteckert2008}
Quantum Monte Carlo approaches based on scattering states\cite{Hershfield1993}
can access  the intermediate coupling regime\cite{HanHeary2007,DirksEtAl2010} at 
finite bias. Recent real-time formulations of
continuous-time quantum Monte Carlo\cite{Rabani08,Werner09,Werner2010}
and an iterative real-time path integral
approach\cite{weissthorwart2008} to quantum transport 
offer the appealing advantage of working directly in the continuum limit, 
but are confined to relatively short time scales.
Access to low temperatures and long times is
hampered in the former case by a severe sign problem,
and by the extrapolation to long memory times in the
latter case. Hence neither approach can presently be
applied to nonequilibrium dynamics of correlated
systems with a small underlying energy scale, as is
the case with ultra-small quantum dots when tuned to
the Kondo regime.

The usage of Lippmann-Schwinger scattering states has 
been well established 
in quantum-field theory\cite{Schweber1962} for over 50 years
and also successfully adapted to the description of 
quantum transport through strongly interacting nano-devices 
coupled to ballistic leads.\cite{Hershfield1993,DoyonAndrei2005,JongHan2006,MethaAndrei2005} 
These states fulfill the correct boundary condition
of the open quantum system: 
(i) they break time-reversal symmetry 
and, therefore, are 
(ii) complex and current-carrying and 
(iii) describe ballistic transport in the leads  combined 
with  scattering events in the small interacting quantum-dot region. 
This time-reversal symmetry breaking is required for current carrying systems 
and reveals itself naturally in all diagrammatic approaches by the occurrence 
of retarded and advanced Green's functions. It is a consequence of any 
regularization when performing the limit to an infinitely large system.

In particular, the work of 
Hershfield \cite{Hershfield1993} and Doyon and Andrei \cite{DoyonAndrei2005} 
has rigorously shown that these boundary conditions remain unaltered 
when a local interaction is switched on. The noninteracting current-carrying 
system evolves 
into the new steady-state of the \emph{interacting} system, and the steady-state density 
operator retains a Boltzmannian form.\cite{Hershfield1993} 
The explicit construction of those scattering states allows to exactly
solve the DC and AC Kondo model at the  
Toulouse point\cite{SchillerHershfield95,SchillerHershfield96,SchillerHershfield2000b}
as well as the interacting resonant level model.\cite{MethaAndrei2005,BoulatSaleurSchmitteckert2008}

Recently, an extension\cite{AndersSSnrg2008} to Wilson's  
numerical renormalization group  has been developed  
for steady-state quantum transport  through nano-devices 
which is able to  deal with the crossover from weak
to strong coupling for arbitrary  bias voltages.
It is based on Oguri's idea\cite{Oguri2007}  of discretizing the 
single-particle scattering states which are the solutions of the Lippman-Schwinger  
equation\cite{Schweber1962}  for the noninteracting problem
and, therefore, fulfill the correct boundary  condition of an open quantum system. 
This scattering-states numerical renormalization group 
approach\cite{AndersSSnrg2008}  (SNRG)  evolves the analytically 
known density matrix  of a noninteracting system 
to the density matrix of the fully interacting problem by employing the 
time-dependent NRG (TD-NRG).\cite{AndersSchiller2005,AndersSchiller2006}
The NRG is ideally suited to the problem, being known to provide 
accurate solutions of quantum-impurity models on all relevant 
interaction strengths at zero bias.\cite{BullaCostiPruschke2007} 
Since the TD-NRG can access exponentially long time scales,\cite{AndersSchiller2005,AndersSchiller2006}
dwell times on the order of the inverse Kondo-temperature are easily accessible.

This paper is organized as follows. After the model used is defined, we
provide the details of the different theoretical approaches in Sec.~\ref{sec:theory}.
We summarize the basic ideas of the SNRG method  
introduced  in Ref.~\onlinecite{AndersSSnrg2008}
in Sec.~\ref{sec:snrg} and state all necessary equations of the
diagrammatic nonequilibrium techniques in Sec.~\ref{sec:KBK-approach}.
The main body of the paper is in Sec.~\ref{sec:results}, where we 
present and discuss the results obtained for the various methods.
In order to set the stage for a detailed comparison between 
the SNRG and diagrammatic  approaches at finite bias, we 
begin with a discussion of the magnetic and charge fluctuation
scales at zero bias in Sec.~\ref{sec:equiScale}. Since the NRG 
provides an accurate solution in this regime for arbitrary 
coupling strengths and temperatures, this reveals the 
validity range of the diagrammatic approaches.
We show that
--- in contrast to the NRG ---
some of the diagrammatic expansions fail to produce 
a single low-energy scale for intermediate and large values of $U$.
However, in the weak correlation regime all these approaches agree excellently for 
arbitrary voltages at small $U$ and yield the same nonequilibrium Green 
functions as well as IV characteristics 
which are presented in Sec.~\ref{sec:weak-coupling}.  
Discrepancies between the different approaches  at intermediate values of the Coulomb 
interaction are discussed in Sec.~\ref{sec:inter-coupling}, where the spectral functions and 
IV curves of a symmetric and an asymmetric junction are considered. 
We conclude with summary and a  short outlook in Sec.~\ref{sec:summary}.

\section{Theory}
\label{sec:theory}

\subsection{Model}
\label{sec:model}

Quantum impurity models are used  to describe quantum transport on the molecular level. Their
Hamiltonian  $\H$
\begin{eqnarray}
  \label{eqn:qis}
  \H &=& \H_{imp} + \H_{bath} + \H_{I}
\end{eqnarray}
consists of three parts: an impurity part  $\H_{imp}$ modeling the interacting device
with a finite number of degrees of freedom, one or several bosonic or fermionic  baths
represented by $\H_{bath}$, and the coupling of these subsystems  by
$\H_I$. 

Throughout this paper, we restrict ourselves to junctions modeled by the single impurity 
Anderson model  with  one
spinful orbital coupled to a left (L) and a right (R) lead and an
on-site Coulomb repulsion $U$
\begin{eqnarray}
  {\H} &=& 
  \sum_{\sigma \alpha=L,R} \int d\e \, (\e -\mu_\alpha) \, c^\dagger_{\e,\sigma\alpha}c_{\e,\sigma\alpha}
  \nonumber \\
&& 
  + \sum_{\sigma = \pm 1}
  E_d \:  \hat{n}^d_\sigma
  + U\hat{n}^d_\uparrow\hat{n}^d_\downarrow
  \nonumber \\
  && + \sum_{\alpha\sigma}   t_{\alpha\sigma} \int d\e \,\sqrt{\rho_\alpha(\e)}
  \left\{
    d^\dagger_\sigma c_{\e\sigma\alpha} +
    c^\dagger_{\e\sigma\alpha} d_\sigma
  \right\} .
\label{eqn:siam}
\end{eqnarray}
Here, $E_d$ is the single-particle energy  of the quantum dot,
$\hat{n}_{\sigma}^{d} = d^{\dagger}_{\sigma} d_{\sigma}$  measures its  
orbital occupancy  and $t_{\alpha\s}$ represent the elementary 
hybridization-matrix elements coupling the dot to the two leads.
The different chemical potentials $\mu_\alpha$ in both leads appear as a 
shift of the band centers and are  functions of the external voltage $V=\mu_R -\mu_L$.

For simplicity, we assume that both leads have the same density of states,
$\rho_R(\e)=\rho_L(\e)\equiv\rho(\e)$, 
characterized by the same band width $D$ but different band centers.
This Hamiltonian is commonly used to model a single
Coulomb-blockade resonance in ultra-small quantum
dots.\cite{MeirWingreen1994,NatureGoldhaberGordon1998}

\subsection{Scattering-states numerical renormalization group  approach}
\label{sec:snrg}
\subsubsection{Definition of the scattering states}
\label{sec:scattering-state}

In the absence of the local Coulomb
repulsion  $H_U=  U\hat{n}^d_\uparrow\hat{n}^d_\downarrow$,
 the single-particle problem is diagonalized exactly in the continuum limit\cite{JongHan2006,Hershfield1993,EnssSchoenhammer2005,HanHeary2007,Oguri2007,LebanonSchillerAndersCB2003,AndersSSnrg2008}
by  the following scattering-states creation operators
\begin{eqnarray}
  \label{eq:scattering-states-operators}
  \gamma^\dagger_{\e \sigma \alpha} &=& c^\dagger_{\e \sigma \alpha} +
  t_\alpha \sqrt{\rho_\alpha(\e)} G_{0\sigma}^r(\e)
 \Bigg[
 d^\dagger_{\sigma} 
 \nonumber \\
 &&
 +
 \sum_{\alpha'} 
 \int d\e' 
 \frac{t_{\alpha'} \sqrt{\rho_{\alpha'}(\e')}}{\e+i\delta -\e'}
 c^\dagger_{\e'\sigma\alpha'} 
 \Bigg]
 \punkt
\end{eqnarray}
$\alpha=L (R)$ labels left (right) moving scattering states created by $\gamma^\dagger_{\e \sigma L(R)}$.
The local retarded resonant level Green's function 
\begin{eqnarray}
G^r_{0\sigma}(\w) &=& \left[\w+i\delta- E_d  - 
\sum_\alpha t^2_\alpha \int d\e \frac{\rho_\alpha(\e)}{\w+i\delta-\e} \right]^{-1}
\end{eqnarray}
enters as an expansion coefficient. 
Defining $\bar t = \sqrt{t_L^2 +t_R^2}$, we will use $r_{R(L)} = t_{R(L)}/\bar t $ and
\begin{eqnarray}
\label{eq:DeltaDef}
\Delta(\w) &=& \bar t^2 \sum_\alpha r^2_\alpha \int d\e \frac{\rho_\alpha(\e)}{\w+i\delta-\e}
\nonumber \\
& =& \Re e[\Delta(\w)] - i\Gamma(\w)
\end{eqnarray}
in the following.

In the limit of infinitely large leads --- volume $Vol.\to\infty$ --- the single-particle spectrum
remains unaltered, and these scattering states diagonalize  the
Hamiltonian
(\ref{eqn:siam})  for $U=0$:
\begin{eqnarray}
  \H^i_0 = \H(U=0) & =& \sum_{\alpha=L,R;\sigma} \int d\e \, \e 
  \gamma^\dagger_{\e\sigma\alpha}\gamma_{\e\sigma\alpha}
  \punkt
  \label{eq:h0-U0}
\end{eqnarray}

The scattering states are solutions of the Lippmann-Schwinger equation\cite{Schweber1962}
and therefore break time-reversal symmetry, which constitutes a  necessary  boundary condition
to describe a current carrying open quantum system. 
This is encoded in the small imaginary part $+i\delta$ entering 
Eq.~(\ref{eq:scattering-states-operators}) -- (\ref{eq:DeltaDef})
required for convergence when performing the continuum 
limit $Vol.\to\infty$ in the leads.

The complex expansion coefficients in (\ref{eq:scattering-states-operators}) are
given by \textit{retarded} functions, e.g.\ $G_{0\sigma}^r(\e)$, which causes 
the scattering states to be  
complex and current carrying. 
For zero bias voltage, time-reversal symmetry manifests itself in the identical
spectrum for left and right movers which are time-reversal pairs in that limit.

To avoid any contribution from  bound states, we will implicitly assume a wide 
band limit: $D\gg max\{ |E_d|,\Gamma , |V|\}$,  where $\Gamma_\alpha= \pi t_\alpha^2\rho(0)$ 
and $\Gamma=\Gamma_L+\Gamma_R$.

Hershfield  has shown that the density operator for such a
noninteracting current carrying quantum system 
retains its Boltzmannian form\cite{Hershfield1993}
\begin{eqnarray}
  \hat \rho_0 &= &\frac{e^{-\beta(\H^i_0 -\hat Y_0)}}{\Tr{e^{-\beta(\H^i_0 -\hat Y_0)}}}
 \label{eqn:rho_0}
\, , \,
 \hat Y_0 = \sum_{\alpha\sigma} \mu_\alpha \int d\e \,
 \gamma^\dagger_{\e\sigma\alpha}\gamma_{\e\sigma\alpha} 
\end{eqnarray}
even for finite bias. The $\hat Y_0$ operator accounts for the
different occupation of the left- and right-moving scattering states, and
$\mu_\alpha$ for the different chemical potentials of the leads.  

Therefore, all steady-state expectation values of operators can be calculated 
using $\hat \rho_0$ which includes the finite bias.  In the absence of a Coulomb 
repulsion $U$, this is a trivial and well-understood problem. It was 
shown\cite{Oguri2007}  that the current expectation value using this 
density-operator $\hat\rho_0$ reproduced the standard 
result\cite{HershfieldDaviesWilkins1991,MeirWingreen1992,MeirWingreen1994} for 
noninteracting devices. The knowledge of the analytical form of $\hat \rho_0$, 
however,  makes this steady-state model accessible  to a NRG  approach.\cite{BullaCostiPruschke2007,AndersSSnrg2008}

The expansion coefficients of $\gamma^\dagger_{\e \sigma \alpha} $ in 
Eq.~(\ref{eq:scattering-states-operators}) contain the complex single-particle Green 
function $ G_{0\sigma}^r(\e)$ which we separate in modulus and phase
\begin{eqnarray}
 G_{0\sigma}^r(\e) &=& | G_{0\sigma}^r(\e)| e^{-i\Phi_\sigma(\e)}
 \quad .
\end{eqnarray}
This phase is absorbed into the  new scattering states  
$ \gamma^\dagger_{\e \sigma \alpha} \to  \tilde \gamma^\dagger_{\e \sigma \alpha}  
=  \gamma^\dagger_{\e \sigma \alpha}  e^{i\Phi_\sigma(\e)}$ by a local gauge transformation. The 
impurity operator $d^\dagger_\sigma$ is expanded into left- and right-mover contributions 
\begin{eqnarray}
\label{eqn:d-LR-expand}
d^\dagger_\sigma &=& r_R d^\dagger_{\sigma R} + r_L d^\dagger_{\sigma L}
\end{eqnarray}
using the inversion of  Eq. (\ref{eq:scattering-states-operators}). These two new operators 
$d^\dagger_{\sigma\alpha}$ are then defined as
\begin{eqnarray}
\label{eqn:d-alpha-cont}
 d^\dagger_{\sigma \alpha} &=& \bar t \int d\e \sqrt{\rho (\e)} | G_{0\sigma}^r(\e )| \tilde \gamma ^\dagger_{\e \sigma \alpha}
 \;\; ,
\end{eqnarray}
and obey the anti-commutator relation $\{ d_{\sigma\alpha},  d_{\sigma'\alpha'}^\dagger   \} =\delta_{\alpha\alpha'}\delta_{\sigma\sigma'}$.

\subsubsection{Discretization of the scattering states}

The scattering-states numerical renormalization group approach\cite{AndersSSnrg2008} (SNRG)
starts from a
logarithmic discretization of the scattering-states continuum $\gamma_{\e\sigma\alpha} $
in intervals  
$I^n_+=[\Lambda^{-(n+z)}D,\Lambda^{-(n+z-1)}D]$ and $I^n_-=[-\Lambda^{-(n+z-1))}D,-\Lambda^{-(n+z)}D]$
$(n=1,2,\cdots )$, controlled by the parameters\cite{Wilson75,BullaCostiPruschke2007} $\Lambda > 1$
and  $z\in (0,1]$. 
The intervals for $n=0$ are defined as $I^0_+=[\Lambda^{-z}D,D]$ and $I^0_-=[-D,-\Lambda^{-z}D]$. 
An average over various $z$-values\cite{YoshidaWithakerOliveira1990} is used to mimic 
the conduction band continuum.

Then, the discretized version of the noninteracting Hamiltonian (\ref{eq:h0-U0}) is mapped onto a
semi-infinite Wilson chain
\begin{widetext}
\begin{eqnarray}
\label{eqn:H0-lambda}
H_0(\Lambda) &=& 
\sum_{\sigma\alpha}\sum_{n=0}^\infty w_{n\sigma\alpha} f^\dagger_{n\sigma\alpha} f_{n\sigma\alpha}
+
\sum_{\sigma\alpha}\sum_{n=0}^\infty \left( t_{n\sigma\alpha}  f^\dagger_{n\sigma\alpha} f_{n+1\sigma\alpha}
+
 t^*_{n\sigma\alpha}  f^\dagger_{n+1\sigma\alpha} f_{n\sigma\alpha}
 \right)
\end{eqnarray}
\end{widetext}
whose tight-binding matrix elements $t_{n\sigma\alpha}$ decay exponentially $t_{n\sigma\alpha}\propto \Lambda^{-n/2}$
for large $n$. 
In contrast to the standard NRG,\cite{Wilson75,BullaCostiPruschke2007} the impurity degree of freedom 
has been included into $H_0(\Lambda)$ since not the leads but the full scattering states have been 
discretized.  Any complex phase in the tight-binding parameters $t_{n\sigma\alpha}$ can be absorbed into the 
creation (anihilation) operators $f^\dagger_{n\sigma\alpha}$ $(f_{n\sigma\alpha} )$ of an electron on the
chain link $n$ with spin $\sigma$ and mover $\alpha$ by a local gauge transformation.

We use $d_{\sigma\alpha}$  defined in Eq.~(\ref{eqn:d-alpha-cont}) as starting vector 
$f_{0\sigma\alpha}=d_{\sigma\alpha}$ for the Householder transformation\cite{Wilson75} and obtain the 
tight-binding coefficients of the Wilson chain (\ref{eqn:H0-lambda})
by the usual procedure.\cite{Wilson75,BullaCostiPruschke2007}  It is straight forward to 
shown that the energy of the first chain link corresponds to the energy of the original quantum-dot 
orbital: $w_{0\sigma\alpha}= E_d$.

\subsubsection{Local Coulomb interaction}

In order to include the local Coulomb interaction, the 
density operator $\hat n^d_\sigma = d^\dagger_{\sigma}d_{\sigma}$
must be expanded  in the new orbitals $d_{\sigma\alpha}$. It consist of  two contributions: 
A density term and a backscattering term $\hat n^d_\sigma =  \hat{n}^0_\sigma +\hat O_{\sigma}^{back}$, where
\begin{eqnarray}
\hat{n}^0_\sigma &=& \sum_\alpha r_\alpha^2
d^\dagger_{\sigma\alpha}d_{\sigma\alpha}
\end{eqnarray}
and the backscattering $\hat O_{\sigma}^{back}$ term is defined as
\begin{eqnarray}
\label{eq:back-scattering}
  \hat O_{\sigma}^{back} &=& r_L r_R\left( d^\dagger_{\sigma R}
    d_{\sigma L}
+d^\dagger_{\sigma L} d_{\sigma R}
\right)
.
\end{eqnarray}
The  local Coulomb interaction $H_U$
\begin{eqnarray}
H_U &=&
U 
\left(\hat n^0_\uparrow \hat n^0_\downarrow
+
\sum_\sigma  \hat O_{\sigma}^{back}  \hat n^0_{-\sigma}
+
 \hat O_{\uparrow}^{back} \hat O_{\downarrow}^{back}
\right)
\end{eqnarray}
leads to a mixing of left and right movers since 
 $\hat O_{\sigma}^{back}$ does not commute with  $Y_0$. However,  the term $H_U^0$,
 \begin{eqnarray}
H_U^0 = \frac{U}{2}
\left( \sum_{\sigma} \hat{n}^0_\sigma -1\right)^2
,
\end{eqnarray}
commutes with $\hat Y_0$ and can be absorbed into the steady-state density operator
$\hat \rho_0\to \tilde \rho_0=\exp[-\beta(\H^i-\hat Y_0)]/Z$ with $\H^i = \H^i_0 +  H_U^0$ 
using the arguments given in Ref.~\onlinecite{DoyonAndrei2005}.

\subsubsection{Review of  the time-dependent numerical renormalization group approach}
\label{sec:td-nrg}

Starting from an equilibrated system for times $t\le 0$, the initial Hamiltonian  ${\H_i}$ is 
changed to ${\H_f}$ by a sudden quench at $t=0$. Then, the density operator $\hat \rho(t)$ evolves  
from its initial value $\hat \rho_0$ at $t=0$ as
\begin{eqnarray}
\hat \rho(t) &=& e^{-i\H_f t/\hbar}\hat  \rho_0 e^{i\H_f t/\hbar}
\punkt
\end{eqnarray}
If $\H_{i(f)}$ describes a quantum impurity problem and $\hat O$ is an impurity operator, it was 
recently shown that 
the real-time dynamics of the expectation value of $O(t) = \expect{\hat O (t)}$ can be 
calculated\cite{AndersSchiller2005,AndersSchiller2006} by evaluating
\begin{eqnarray}
\label{eqn:o-vs-t-td-nrg}
O(t) &=& \sum_{m}\sum_{r,s}^{dis}  \rho^{red}_{r,s}(m)  O^m_{s,r}\, e^{-i(E^m_r -E_s^m)t/\hbar} 
\end{eqnarray}
where $ O^m_{s,r}=\bra{s,e;m}\hat O \ket{r,e;m}$ denotes the matrix elements of 
the operator $\hat O$ and $E^m_r$ the NRG eigenenergy of the eigenstate $\ket{r;m}$ to $\H_f$ 
at NRG iteration $m$. 
The sum restriction $\sum_{r,s}^{dis} $ indicates that at least one of the 
states $r,s$ must be a discard state at iteration $m$. Excitations between two 
retained states will be refined in the following iterations $m'>m$ 
and, therefore, will contribute at a later iteration. $e$ labels the environment degrees of freedom 
of the Wilson chain links to be incorporated in subsequent iterations $m'>m$
and the  $\ket{r,e;m}= \ket{r;m}\otimes\ket{e}$ are just tensor-product 
states of the eigenstates of the $m$th iteration and the yet uncoupled rest chain.
The reduced density matrix
\begin{eqnarray}
\label{eqn:red-dm}
\rho^{red}_{r,s}(m)  &=& \sum_{e} \bra{r,e;m}\hat \rho_0\ket{s,e;m} 
\end{eqnarray}
traces out all environment degrees of freedom $e$. The initial conditions are encoded into the density
operator $\hat \rho_0$ calculated with the initial Hamiltonian  ${\H_i}$. The calculation of the overlap matrix
between the NRG eigenstates of ${\H_i}$ and ${\H_f}$ allows for the basis set transformation 
of $\rho^{red}_{r,s}(m)$ into the basis of the final Hamiltonian provided that $\hat \rho_0$ remains restricted to
the last Wilson shell.\cite{AndersSchiller2005,AndersSchiller2006} This transformed  $\rho^{red}_{r,s}(m)$ 
enters Eq.~(\ref{eqn:o-vs-t-td-nrg}).

The discarded states form 
a complete basis set\cite{AndersSchiller2005} for the Fock-space of the \textit{entire} Wilson chain
of length $N$, i.e.\ $\mathcal{F}_N=\mathrm{span}\{\ket{l,e;m}\}$ where $l$ labels all discarded
states at iteration $m$.
The iterative diagonalization thus procures the set of (approximate) eigenstates  for the \textit{whole} 
energy range from high energies on the order of the bandwidth down to very low energies such as
the Kondo scale.  This is indispensable because nonequilibrium processes usually  
involve all energy scales and cannot be confined to a finite low energy window  
set by the last Wilson shell as in the usual equilibrium NRG.

\subsubsection{The scattering-states NRG approach and steady-state Green's function}

In Sec.~\ref{sec:scattering-state} we have argued that the analytic form the  steady-state 
nonequilibrium density operator is known for the noninteracting case. This  allows for 
applying the NRG approach to construct a faithful representation of  $\hat\rho_0(V,U=0)$. 
We assume that when switching on the Coulomb interaction $H_U$ for infinitely large leads  
(i) a  steady state is reached after some characteristic 
but finite time and (ii) it is unique and independent of the initial condition. 
As described earlier, the boundary condition of time-reversal symmetry breaking
is imposed on the scattering states and the nonequilibrium density operator at $t=0$ for $U=0$.
The interaction quench at $t=0$, i.e.\ switching on a local scattering potential, 
and the subsequent unitary  time evolution do not affect this boundary condition, and 
the time-evolved operators characterize the interacting current carrying open quantum system.

The time average of  the density operator
\begin{eqnarray}
\hat \rho_\infty &=& \lim_{T\to\infty} \frac{1}{T}\int_0^T dt \hat \rho(t)
\end{eqnarray}
projects out the steady-state contributions to the time-evolved 
density operator $\hat \rho(t)=\exp(-i\H_f t/\hbar)\hat \rho_0\exp(i\H_f t/\hbar)$ 
even in a finite-size system: only the energy diagonal terms contribute in accordance 
with the steady-state condition $[\H_f,\hat \rho_{\infty}] =0$. Even though $\hat \rho_{\infty}$ remains unknown 
analytically, we can construct it systematically using the 
time-dependent NRG\cite{AndersSchiller2005,AndersSchiller2006} described above. 

The steady-state retarded Green's function is defined as
\begin{eqnarray} 
\label{eq:neqGf}  G^r_{A,B}(t) &= & - i \Tr{ \hat \rho_{\infty} [ \hat A(t), \hat B ]_s } \Theta(t), \label{eqn:steady-state-gf}   
\end{eqnarray}
where $\hat A(t)=e^{i\H_f t/\hbar}\hat A e^{-i\H_f t/\hbar}$,
$[ \hat A(t), \hat B ]_s$ denotes the commutator ($s=-1$) for bosonic, and the 
anti-commutator ($s=1$) for fermionic correlation functions.
This Green's function can be calculated using the time-dependent NRG\cite{AndersSchiller2005,AndersSchiller2006} 
and extending ideas developed for equilibrium Green's functions.\cite{PetersPruschkeAnders2006}
The completeness relation 
for the basis of discarded states as introduced above is given by  
\begin{eqnarray} 
\label{eqn:completness-nrg}
1&=&\sum_{m=m_{min}}^{N}\sum_{l\in dis}\sum_{e}\ket{l,e;m}\bra{l,e;m}\quad,
\end{eqnarray} 
where $m_{min}$ denotes the first iteration at which the NRG truncation is employed, $N$ is the total number 
of iterations (i.e.\ the length of the Wilson chain), 
and $l$ only runs over the states which are discarded at iteration $m$. 
For each iteration $m$, we can partition the completeness relation (\ref{eqn:completness-nrg})
into two parts, $1=1^-_{m}+1^+_{m}$, where the first part incorporates the iterations $m_{min}$ to $m$ and the 
second the iterations $m+1$ to $N$.  Since $1^+_{m}$ spans the part of the Fock-space which contains 
all kept states  $\ket{k,e;m}$ after iteration $m$, the identity
\begin{eqnarray}
\label{eqn:kept-states}
1^+_{m} &=& \sum_{m'=m+1}^{N}\sum_{l\in dis}\sum_{e}\ket{l,e;m'}\bra{l,e;m'}\nonumber \\
&=& \sum_{k\in kept} \sum_{e}\ket{k,e;m}\bra{k,e;m}
\end{eqnarray}
must hold. The different contributions to the  Green's function are calculated for each energy scale 
$D_m\propto \Lambda^{-m/2}$ by expanding the (anti-)commutator in Eq.~(\ref{eq:neqGf}) and inserting the 
completeness relations Eqs.~(\ref{eqn:completness-nrg}) and (\ref{eqn:kept-states}) repeatedly.
By making use of the fact that local operators $\hat A$ and $\hat B$ are
diagonal in the environment degree of freedom $e$, reduced density matrices $\rho^{red}_{r,s}(m)$ occur
naturally when tracing out the environment  $e$ here as well.
Although the excitation energies remain confined to the same energy scale, terms
connecting different energy scales $D_m$ and $D_{m'}$ are implicitly included 
through the reduced density matrices such as 
defined in (\ref{eqn:red-dm}). Similar to the real-time dynamics,
the summation over all $m$ then ensures that all energy scales $D_m$
contribute to the Green's functions.\cite{PetersPruschkeAnders2006,AndersNeqGf2008} 
A detailed  derivation is given in Ref.~\onlinecite{AndersNeqGf2008}.
It was shown that the algorithm is identical to the equilibrium 
algorithm\cite{PetersPruschkeAnders2006} if $\H_i=\H_f$. 
Laplace transforming  $G^r_{A,B}(t)$ 
yields the steady-state spectral function for the retarded Green's function 
which is used to calculate the current (see Eq.~(\ref{eq:ss-current}) below).

\subsection{The Kadanoff-Baym-Keldysh approach}
\label{sec:KBK-approach}

\begin{figure*}[tbp]
  \includegraphics[scale=1]{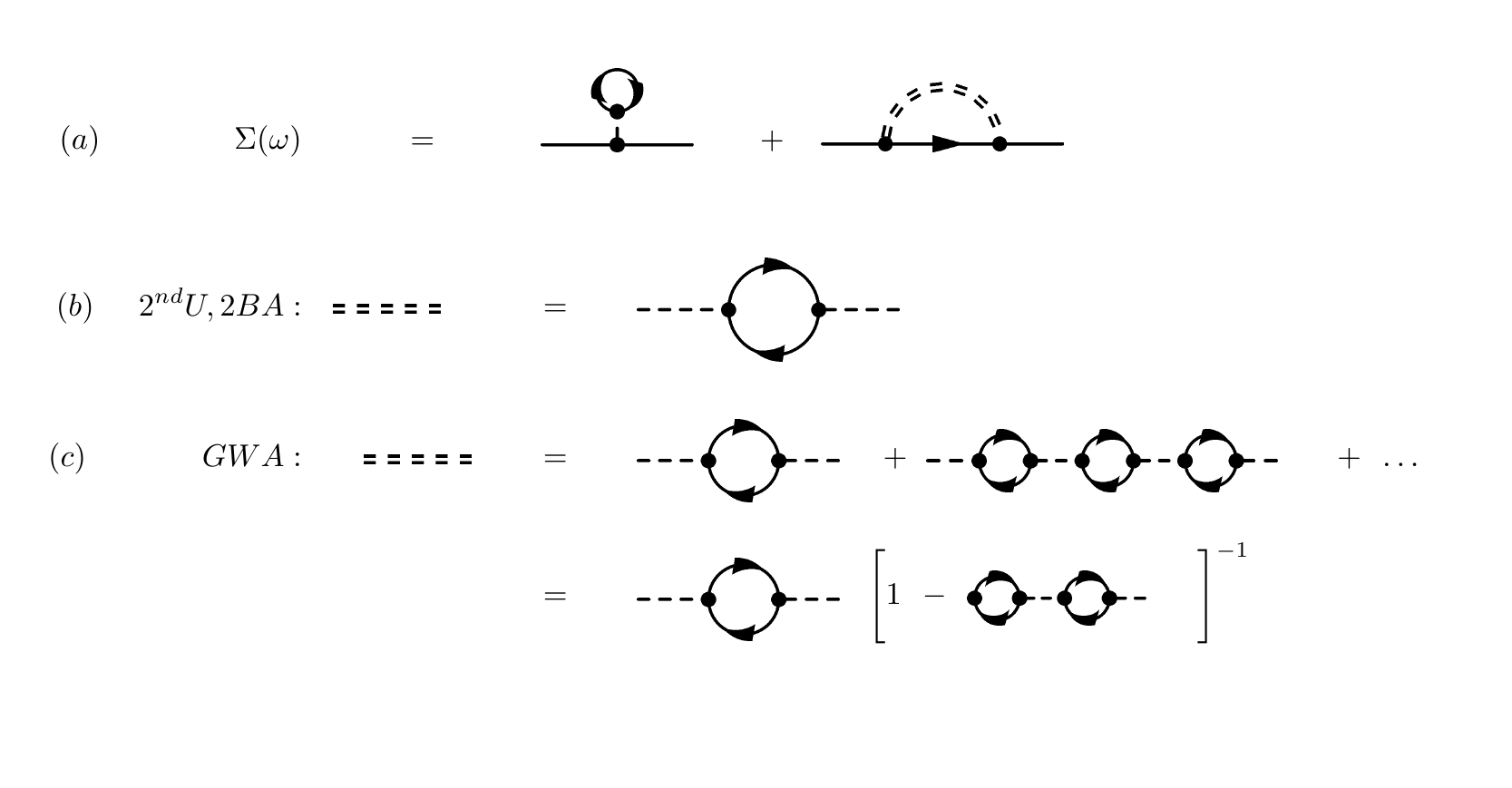}
  \caption{(a) Schematic diagrammatic representation of the Kadanoff-Baym-Keldysh self-energy. 
    The first term represents the frequency independent Hartree shift while the second 
    contribution represents the interaction part. The double-dashed line is the effective 
    interaction $W$ while the single-dashed line represents the bare interaction $\sim U$. 
    The second-order diagram is shown in (b). For the non self-consistent second-order $U$ 
    ($2^{nd} U$) approximation the internal solid lines with arrows are taken as the 
    Hartree-Fock (HF) propagators, 
    while for the  second Born approximation (2BA) the internal lines denote fully-dressed 
    propagators.
    In the GW approximation (GWA) (c) the interaction is renormalized by an infinite series of particle-hole 
    pairs which can be summed as indicated in the last line. The internal lines again denote fully-dressed 
    propagators.
  }
  \label{fig:perturb}
\end{figure*}

We employ the nonequilibrium perturbation theory as formulated by Kadanoff and Baym\cite{kadanoffBook62} 
and Keldysh \cite{Keldysh65} on the usual Keldysh time contour, for
example, see Refs.~\onlinecite{danielewiczQFTNonEqI84,wagnerNeqGF91,leeuwenKeldyshIntro06}.
Since we are only interested in the steady-state properties, the information and correlations
of the initial conditions are assumed to be lost. This is archived by sending the
initial time  $t_0\to-\infty$  and dropping all
correlation functions which involve the initial state. It is again assumed, that the system 
reaches a steady state  which is translational invariant in time. Therefore,
the single-particle Green's function does only depend on the difference 
between the two formerly independent times of particle creation and annihilation.
The Laplace transform of the time difference then leads to the formulation in
frequency space for all Green's functions of the steady state.     

In the nonequilibrium steady-state formulation two independent components  of the contour 
ordered Green's function survive which are chosen to be the retarded and lesser Green's functions,
$G^r(\w)$ and $G^<(\w)$ respectively. The advanced and greater functions are related via
\begin{eqnarray}
  \label{eq:advGF}
  G^a(\w)&=&G^r(\w)^\dagger
  \\
  \label{eq:grtGF}
  G^>(\w)&=&G^<(\w)+ G^r(\w)-G^a(\w)
\quad.
\end{eqnarray}

The two relevant Green's functions can be expressed as \cite{haugBookTransport96,spataruGwMillis2009}
\begin{eqnarray}
  \label{eq:Grdef}
  G^r_\s(\w)&=&\frac{1}{\w+i\delta-E_d-\Sigma^H_\s-\Delta(\w)-\Sigma^r_\s(\w)}
  \\ \nonumber &\\
  \label{eq:GLessdef}
  G^<_\s(\w)&=&|G^r_\s(\w)|^2 \Big[2i \feff(\w)+\Sigma_\s^<(\w)\Big]
  \\
   \feff(\w) &=&  f_L(\w)\Gamma_L(\w)+f_R(\w)\Gamma_R(\w)
\label{eqn:f-eff}
  \quad,
\end{eqnarray}
where, again, $\Delta(\w)=\Delta_R(\w)+\Delta_L(\w)$ are the hybridization functions of the leads, 
$\Gamma_\alpha(\w)$ their imaginary parts (see Eq.\ (\ref{eq:DeltaDef}))
and $f_\alpha(\w)=1/\{\exp\beta(\w-\mu_\alpha)+1\}$ are the Fermi functions 
of the corresponding leads. 
The retarded and lesser self-energies, $\Sigma^r$ and $\Sigma^<$ respectively, include all 
correlation effects induced by the Coulomb interaction $U$.
$\Sigma^H_\s$ accounts for the  frequency independent 
Hartree energy shift.

\subsubsection{Nonequilibrium self-energy}

Diagrammatic expansions in the Coulomb interaction of the self-energy \cite{luttinger:LuttWardII60} 
have been investigated for systems in equilibrium \cite{yamadaSIAMPerturbIV75,horvaticSIAM82,chu:parquetSIAM92,whiteSCSOPT92}
as well as in nonequilibrium.\cite{HershfieldDaviesWilkins1991,hershfieldTunnelingSIAM92,yeyatiNonEqMPT93,takagiMagFieldSIAM99,matsumotoNeqSIAM00,fujiiPerturbKondoNeq03,thygesenNonEqGW07,darancetNeqGW07,thygesenNonEqGW08,spataruGwMillis2009}
The self-energies can 
be evaluated either non self-consistently, where bare propagators
are used as inner lines, or in terms of skeleton diagrams, where fully-dressed propagators are 
taken into account.

In this study we focus on three different approximations for the self-energy: (A) The bare 
expansion up to second order in $U$, where Hartree-Fock (HF) propagators are used as 
internal lines. The latter are just the noninteracting propagators, but with a shifted
level position $E_d'=E_d+U/2$.
This approximation is labeled $2^{nd} U$ and its diagrammatic representation is schematically shown
in Fig.~\ref{fig:perturb}(a) and (b).  (B) The self-consistent evaluation
of the second-order skeleton diagram of Fig.~\ref{fig:perturb}(a) and (b).
This approximation is  called second Born approximation 
(2BA) but in contrast to the usual 2BA, no exchange contribution exists for the single 
impurity Anderson model~(\ref{eqn:siam}) with only one spinful orbital.
(C) In the GW approximation\cite{AryasetiawanGW98,aulburGW99} 
(GWA) the bare Coulomb interaction $U$ is screened by an infinite series 
of particle-hole excitations, which can be summed as indicated in Fig.~\ref{fig:perturb}(c).
No contributions with odd orders in the interaction occur in this
series due to the definition of the matrix elements of the Coulomb interaction in our model~(\ref{eqn:siam}),
where we set matrix elements between electrons with the same spin explicitly to zero.\footnote{For a 
discussion of a different definition see the appendix of
Ref.~\onlinecite{wangGWSIAM07}. } 

The 2BA and the GWA are both evaluated self-consistently, and thus the self-energies can 
be derived from a Luttinger-Ward functional.\cite{luttinger:LuttWardII60} Therefore, both constitute 
conserving approximations in the sense of Kadanoff and Baym.\cite{baym:conservingApprx62} It can be 
shown that elementary sum rules such as charge and current conservation are obeyed.\cite{thygesenNonEqGW08}
In contrast, the non self-consistent $2^{nd}U$ approximation is not conserving
which can lead to the violation of current conservation, as it will be demonstrated later.

The Hartree shift is produced by the average occupation of the quantum dot 
\begin{eqnarray}
  \label{eq:SigHartree}
  \Sigma^H_\s&=&U\ew{\hat n_{\bar \s}}  \\
  \label{eq:occupation}
  \ew{\hat n_{\s}}&=&  \int\frac{d\w}{2\pi i}G^<_{\s}(\w)
\end{eqnarray}
and analytic expressions for the self-energies read
\begin{eqnarray}
  \label{eq:SigR}
  \Sigma^r_\s(\w)&=&
  i \int\frac{dx}{2\pi} G^<_\s(x)\:W^r_\s(\w-x)
  \non
  &&
  +
  i \int\frac{dx}{2\pi} G^r_\s(x)\:W^>_\s(\w-x)
  \\\label{eq:SigLess}
  \Sigma^<_\s(\w)&=&i \int\frac{dx}{2\pi} G^<_\s(x)\:W^<_\s(\w-x)
  \quad,
\end{eqnarray}
where the effective interactions are given by 
\begin{eqnarray}
  \newcounter{HlpCounter}
  \setcounter{HlpCounter}{\value{equation}}
  \label{eq:WrBA}
  W_\s^r(\w)&=U^2 P^r_{\bar\s}(\w) \qquad&\qquad (2BA)
  \\
  \label{eq:WrGW}
  W_\s^r(\w)&=\frac{U^2 P^r_{\bar\s}(\w)}{1-U^2 P^r_{\s}(\w)\:P^r_{\bar\s}(\w)} \qquad
&\qquad (GWA)  \\
  \label{eq:WLess}
  W_\s^<(\w)&=W^r_\s(\w)\:P_\s^<(\w)\:W^a_\s(\w)&
  \\
\label{eq:WGtr}
  W_\s^>(\w)&=W^r_\s(\w)\:P_\s^>(\w)\:W^a_\s(\w)&
  ,
\end{eqnarray}
and  the particle-hole bubbles are 
\begin{eqnarray}
  \label{eq:Pr}
  P_\s^r(\w)&=&-i \int\frac{dx}{2\pi}\: G^r_{\s}(x)\:G^<_{\s}(x-\w)
  \non &&
  -i  \int\frac{dx}{2\pi} \:G^{<}_{\s}(x)\:G^{a}_{\s}(x-\w)
  \\
  \label{eq:Pa}
  P_\s^a(\w)&=&-i \int\frac{dx}{2\pi}\: G^a_{\s}(x)\:G^<_{\s}(x-\w)
  \non
  &&
  -i \int\frac{dx}{2\pi} \:G^{<}_{\s}(x)\:G^{r}_{\s}(x-\w)
  \\
  \label{eq:PLess}
  P_\s^<(\w)&=&-i \int\frac{dx}{2\pi}\: G^<_{\s}(x)\:G^>_{\s}(x-\w)
  \quad.
\end{eqnarray}
In the above expressions the advanced and greater Green's functions can be determined 
via Eq.\ (\ref{eq:advGF}) and (\ref{eq:grtGF}) and $\bar\s=-\s$ denotes the opposite spin of $\s$.

Equations (\ref{eq:Grdef})-(\ref{eq:PLess}) form a closed set, which is solved 
self-consistently 
for the 2BA and GWA. For the $2^{nd}U$ approximation 
all particle-hole propagators (\ref{eq:Pr})-(\ref{eq:PLess}) are evaluated only once 
with bare Green's functions
\begin{eqnarray}
  \label{eq:GFU0}
  g^r_\s(\w)&=&\frac{1}{\w+i\delta-E_d-\Sigma^H_\s-\Delta(\w)}
  \\ \nonumber &\\
  g^<_\s(\w)&=&  2i |g^r_\s(\w)|^2 \feff(\w)
  \quad.
\end{eqnarray}
and Eq.\ (\ref{eq:WrBA}) is used as the effective interaction.
The Hartree shift is included in order to determine the desired filling. 
The effective Fermi function $\feff(\w)$ was defined in
Eq.~(\ref{eqn:f-eff}).

The GWA \cite{AryasetiawanGW98,aulburGW99} has been successfully applied to
overcome some shortcomings of local-density calculations and estimate 
the screening of the Coulomb interactions in solid state physics.
Recently, it has been employed to calculate quantum  transport 
through nanoscale devices.\cite{thygesenNonEqGW07,thygesenNonEqGW08,darancetNeqGW07,spataruGwMillis2009,GWvanLeeuwen2009} 
In the context of the single impurity Anderson model it was shown 
to accurately describe the equilibrium properties in the weakly-interacting regime and in asymmetric
situations with a nearly empty or nearly full impurity orbital.\cite{wangGWSIAM07,thygesenNonEqGW08}
In the strongly interacting Kondo regime, i.e.\ $\Gamma-U<E_d<-\Gamma$, the GWA produces 
a  narrow peak in the spectral function at the Fermi level, which could
be interpreted as remnants of the expected many-body resonance.\cite{thygesenNonEqGW08}
However, the line shape of this low-energy resonance as well as the high-energy Hubbard 
peaks at $\omega\approx E_d$ and $\omega\approx E_d+U$
are not correctly reproduced by this approximation.\cite{wangGWSIAM07,thygesenNonEqGW08}
Additionally, for very large interactions strength
$U/\Gamma>8$, all three perturbative approaches  favor an unphysical magnetic ground state,
which is actually forbidden by the Mermin-Wagner theorem.\cite{merminWagnerTheorem66}
In the nonequilibrium situation, the proximity to bifurcation points of these sets of 
equations leads to unphysical hysteretic response.\cite{spataruGwMillis2009}

\subsection{Current as function of the bias voltage}

The current flowing from lead $\alpha$ onto the impurity region can be expressed as \cite{MeirWingreen1992} 
\begin{eqnarray}
  \label{eq:currentMeirWingreen}
  I_\alpha&=& \frac{e}{h}\sum_\s \int \!d\w  \Gamma_\alpha(\w)\:\Big[
    2i G^<_\s(\w,V)
  \\ \nonumber
  &&\phantom{\frac{e}{h}\sum_\s \int \!d\w \Gamma_\alpha(\w)}
   +f_\alpha(\w)4\pi \rho^r_\s(\w,V)  \Big]
\end{eqnarray}
where $\rho^r_\s(\w,V)=-\Im m[G^r(\w,V)]/\pi$ is the frequency and voltage
dependent spectral function of the retarded impurity Green's function. 
Since the steady-state current onto the interacting region from the left 
must be equal to the current leaving to the right lead, i.e.\ $I_L=-I_R=I$, we can 
symmetrize the left and the right currents with a linear 
combination\cite{MeirWingreen1992} and write it as
\begin{eqnarray}
  \label{eq:currentSym-def}
  I&=& r^2_R I_L-r^2_L I_R
  \quad .
\end{eqnarray}
In the wide band limit, $\Gamma_\alpha(\w)\to \Gamma_\alpha=\Gamma_\alpha(0)$ 
and $r_\alpha^2= \Gamma_\alpha/\Gamma$ holds such that the term 
proportional to $G^<_\s(\w)$ drops out of (\ref{eq:currentSym-def}) and we obtain
\begin{eqnarray}
 \label{eq:currentSym}
   \label{eq:ss-current}
  I&=&
  \frac{G_0}{e}
 \sum_\s \int \!d\w  \left[
   f_L(\w)-f_R(\w)
  \right]
 \pi\Gamma \rho^r_\s(\w,V)
  \:.
  \end{eqnarray}
where we have defined $G_0$
\begin{eqnarray}
G_0 &=&    \frac{e^2}{h}\frac{4\Gamma_L\Gamma_R}{\Gamma^2}
\punkt
 \end{eqnarray}
$G_0$ reaches the universal conductance quantum $e^2/h$ for a symmetric 
point-contact junction, 
$\Gamma_L=\Gamma_R$, and is strongly suppressed in the tunneling regime $\Gamma_\alpha \ll \Gamma_{-\alpha}$. 

For the voltage drop across the two contacts of the impurity to the leads we employ a serial resistor model
where the chemical potentials in the leads are given by $\mu_L=-r^2_R V$ and $\mu_R=r^2_L V$.

At zero temperature, the zero bias conductance $G=edI/dV|_{V=0}=G_0\pi\Gamma\sum_\s\rho^r(0)$
is proportional to the spectral function at the Fermi level. In the zero temperature Fermi liquid
and for a symmetric junction $\rho^r_\s(0)=1/(\pi\Gamma)$. The conductance is given by its universal 
value $G=2G_0$ which shows in the slope at zero bias of the IV characteristics, i.e.\ $I e/G_0=2V$.   

We also define a leakage current
\begin{eqnarray}
  \label{eq:currentLeak}
  \Delta I&=&I_L+I_R
  \\ \nonumber &=& 
  \frac{2e}{h}\sum_\s \int \!d\w\:
    \bigg[\Gamma_L(\w)+\Gamma_R(\w)\bigg] i G^<_\s(\w)  
      \\ \nonumber 
&&
+\frac{4e}{h}\sum_\s \int \!d\w\: \feff(\w) \pi \rho^r_\s(\w)
\end{eqnarray}
which must vanish due to current conservation, $I_L=-I_R$, in a physical junction. 
Therefore, deviations from $\Delta I=0$ measures shortcomings of an approximation.

\section{Results}
\label{sec:results}

In this section we compare and discuss the results obtained from the 
different diagrammatic Keldysh approaches  with the SNRG. 
For simplicity, we used symmetric structureless leads characterized by a constant 
density of states
with a half-bandwidth $D=20\Gamma$, i.e.\ 
$\Gamma_\alpha(\w)=\Gamma_\alpha\:\Theta(D-|\w|)$.
The total $\Gamma=\Gamma_L+\Gamma_R$ is used as 
the energy scale: All energies, voltages and temperatures 
are measured in units of $\Gamma=1$ throughout the paper.

For the SNRG a rather large $\Lambda=4$ was chosen, and $N_s=2200$ 
states were retained in each NRG-iteration step. 
$z$-averaging\cite{YoshidaWithakerOliveira1990} with either $N_z=2$ or $N_z=4$ 
different $z$-values was performed, 
and the broadening parameter for the spectral function\cite{BullaCostiVollhardt01} was 
chosen $b=1.3/N_z$.
For large $U$ some unphysical wiggles 
may emerge in the spectral function, as it is explained below.
In principle, these wiggles can be minimized by choosing a smaller $\Lambda$,
incorporating more states or performing the $z$-averaging with a larger number 
of $z$-values.

We did not include an external magnetic field, and no magnetic solutions are encountered 
for the parameter values used in this paper. Therefore,  we will drop the spin-index from now. 
The  two spin components of the spectral functions  and self-energies are identical,
e.g.\ $\rho^r_\s=\rho^r_{\bar\s}\equiv \rho^r$
and $\Sigma^r_\s=\Sigma^r_{\bar\s}\equiv \Sigma^r$ respectively.

Before we apply finite bias voltages, we will
compare the different equilibrium low-energy scales obtained with the diagrammatic approaches 
to NRG results.
While the diagrammatic approach becomes exact only in the weak-coupling limit $U\to 0$, 
the NRG produces the correct scales for all interaction strengths. We will identify 
the validity range of the diagrammatic expansion. In that regime the diagrammatic approach
produces correct results even in nonequilibrium, and we will therefore use it to benchmark 
the SNRG for finite voltages.

\subsection{Equilibrium low-temperature scales}
\label{sec:equiScale}

The single impurity  Anderson model in equilibrium for $T\to 0$ always forms  
a local Fermi liquid.\cite{nozieres:FLKondo74,KrishWilWilson80a,KrishWilWilson80b,hewson:KondoProblem93}
The spectral function for a symmetric junction approaches the zero temperature
limiting value $\rho(\w=0,T=0)=1/(\pi\Gamma)$  in 
accordance with the Friedel sum rule.\cite{Langreth1966,Anders1991}
The Fermi-liquid formation is associated with a characteristic low-energy scale, which is identified 
with the Kondo temperature $T_K$ at large Coulomb repulsions and near half-filling.

The SNRG coincides with the usual NRG\cite{Wilson75,KrishWilWilson80a,BullaCostiPruschke2007} in 
equilibrium,  which accurately
describes the crossover from high to low temperatures and
provides the correct low-energy scale $T_K$ depending exponentially on $U$.\cite{KrishWilWilson80a}
The $2^{nd}U$ approximation, however, predicts 
a low-energy scale which is perturbative in $U$ and too large.\cite{meyerAsymSIAMMPT99}
The GWA does produce a narrow many-body resonance 
in the spectral function at the Fermi level. Extracting a low-energy scale from 
the full width at half maximum (FWHM) for an asymmetric junction ($E_d\neq  -U/2$),
as shown in Fig.~5 of Ref.~\onlinecite{thygesenNonEqGW08}, 
suggests an exponential variation with the ionic level position $E_d$. 
However, the exponent has the wrong prefactor as compared to the exact analytic form.\cite{Tsvelick82,KrishWilWilson80b}

In order to extract the low-energy scale from our model calculations we employ two different methods:
We calculate the temperature dependent zero bias conductance $G=dI/dV|_{V=0}$ and
fit it to a phenomenological form.\cite{goldhaberSET98,CostiHewsonZlatic94}
Since $G$ is directly determined by the spectral function, it is sensitive to the 
amount of spectral weight in the temperature window $-T\apprle \w \apprle T$.
The scale $T_K^{charge}$ extracted in this way
constitutes the energy scale relevant for the zero-bias charge transport in the system.
This procedure yields the same result as the aforementioned  extraction from the FWHM of the
resonance at the Fermi level.

The second way 
utilizes the screening of the  
effective local magnetic moment, $\mu_{eff}^2=T\chi(T)=T dM/dH|_{H=0}$, where $\chi$ is the 
magnetic susceptibility, $M$ the magnetization and $H$ an external magnetic field.
We calculate $M$ for a finite but small external magnetic field $\delta H=10^{-9}\Gamma$,
and extract the susceptibility via the difference quotient.
In the Fermi-liquid regime the effective magnetic moment 
follows an universal curve as function of temperature  from which the
low-energy scale is determined by defining $T_K\chi(T_K)\approx 0.07$.\cite{Wilson75,KrishWilWilson80a}
The resulting $T_K^{mag}$ sets the scale for
magnetic excitations in the system and is  directly linked to  the 
Kondo-screening of the local magnetic moment.

For large values of the Coulomb interaction, the scales $T_K^{mag}$ and $T_K^{charge }$
should coincide (apart from a constant of order one) and vary as $\exp(-\pi U/8\Gamma)$
for a symmetric quantum dot.
For very small values of the Coulomb repulsion $U\ll\Gamma$ both
should approach $\Gamma$. The charge scale $T_K^{charge}$ is expected to be roughly constant 
and on the order of $T_K^{charge}\sim \Gamma$ for $U/\pi\Gamma\apprle 1$ since for such small 
interactions charge fluctuations to and from the leads dominate the physics,
and the spectral function stays very close to its HF form. 
On the other hand,  the magnetic scale is known to decrease
exponentially 
for all $U$.
\cite{okijiSIAMExactSus83,horvaticExactSIAMPerturb85}

\begin{figure}[tbp]
  \includegraphics[scale=0.65]{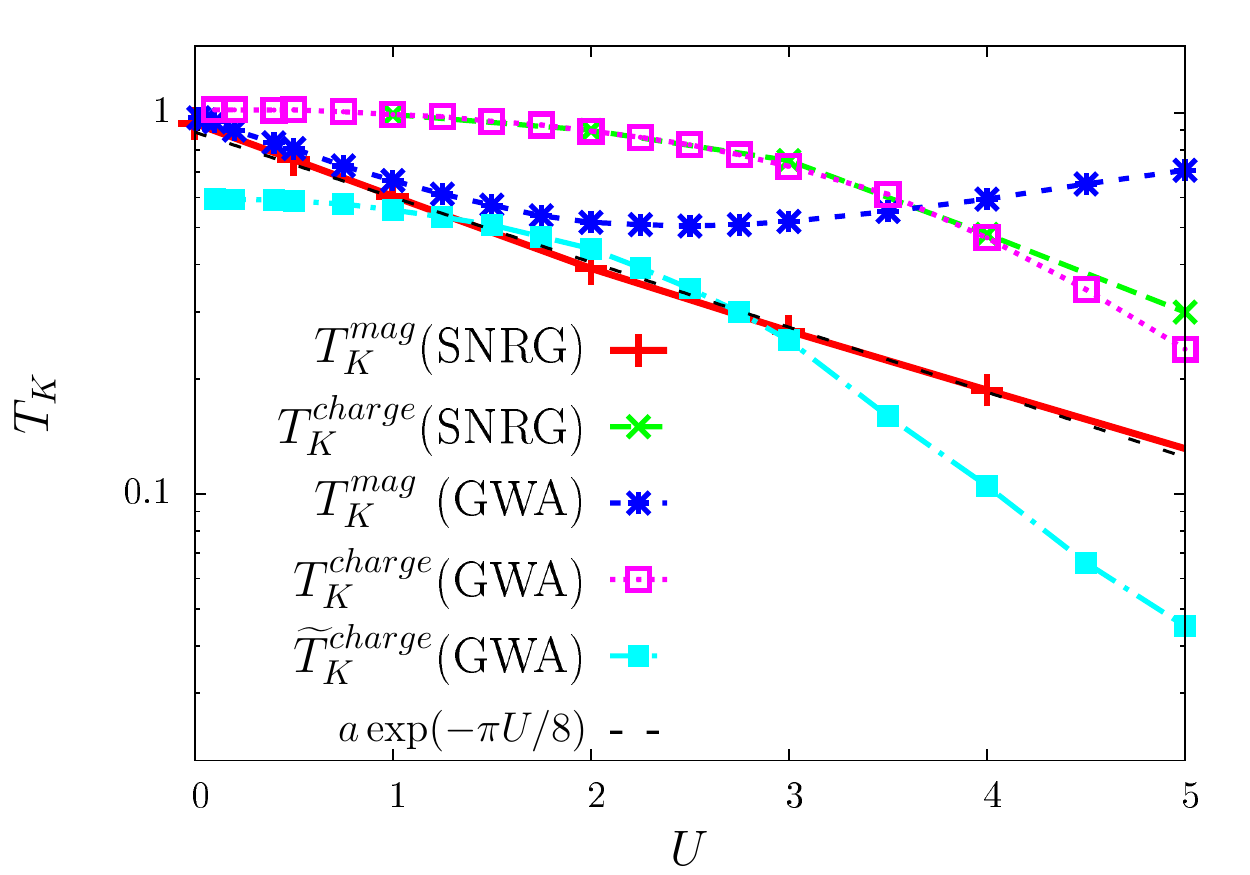}
  \caption{The equilibrium low-energy scales as functions of $U$ extracted from the 
    various approximations as described in the text. A fit to the magnetic scale 
    of the NRG to show the 
    exponential decay $\propto\exp(-\pi U/8)$ is also included in the plot. 
  }
  \label{fig:gfU4-II}
\end{figure}

Figure~\ref{fig:gfU4-II} shows the two scales 
extracted from NRG and GWA calculations for a symmetric junction in equilibrium. 
The NRG results show the expected $U$-dependencies: The charge scale $T_K^{charge}$ is on the order of
$\Gamma$ for small $U\apprle 3\Gamma$ and decreases exponentially for large $U\apprge 4\Gamma$.
The magnetic scale $T_K^{mag}$ decreases exponentially for all $U$
as it is evident from the comparison with a fit function $a \exp(-\pi U/8)$ also included in the plot.
Furthermore, there exists only one universality scale for 
large $U$  which manifests itself by $T_K^{mag}\propto T_K^{charge}$ (not shown). 

On the other hand, the scales obtained from the GWA agree with the NRG 
only for small $U$. The charge scale $T_K^{charge}$ perfectly agrees with
the NRG curve for $U\apprle 4$. Significant deviations are observed for larger $U$,
where  the GWA-$T_K^{charge}$ decreases  faster than  the NRG. 
For $U$  significantly larger than the ones shown in the plot, no scales could 
be extracted due to the artificial symmetry breaking 
already reported in the literature.\cite{wangGWSIAM07,thygesenNonEqGW08}

We added a second GWA charge scale  $\widetilde T_K^{charge}$ to the graph 
which is obtained from the
width of the low-energy feature at $75\%$ of $\rho^r(0)$ 
(and not at the FWHM as for $T_K^{charge}$).
The correspondingly extracted scale should 
coincide with $T_K^{charge}$, apart from a prefactor.
But it is found that both scales follow the same trend only for 
small $U$ and 
already for $U\apprge3$ a  much stronger decrease than the 
expected $\exp(-\pi U/8\Gamma)$ is observed in  $\widetilde T_K^{charge}$ .

\begin{figure}[tbp]
  \includegraphics[scale=0.65]{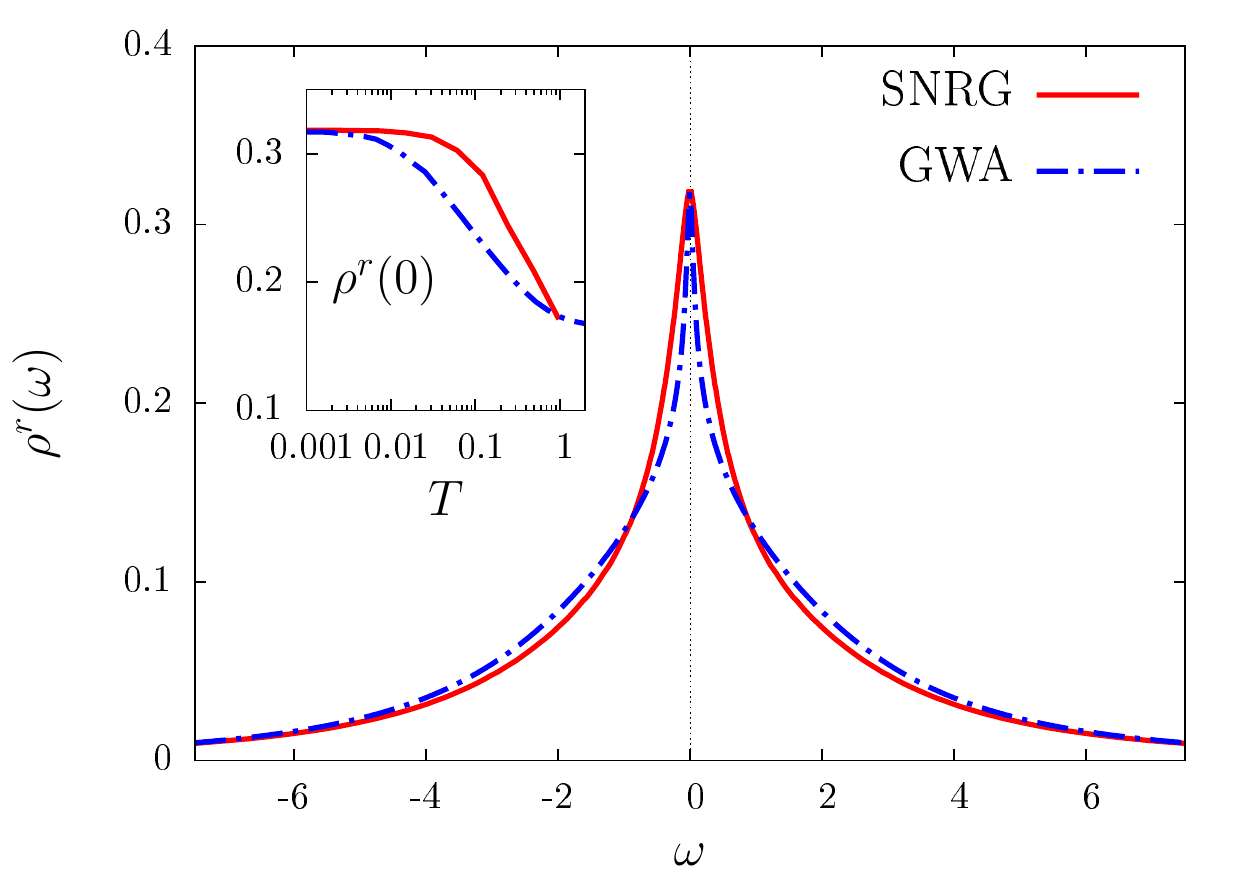}
  \caption{Comparison of the NRG and the Keldysh GWA equilibrium ($V=0$) zero-temperature spectral
    functions for $U/\Gamma=4$ and $E_d=-U/2$ and a quantum-point contact $\Gamma_L=\Gamma_R=1/2$.
    The inset shows the temperature evolution of the spectral function right at the Fermi level, 
    $\rho^r(\w=0,T)$. The NRG-parameters are $\Lambda=2$, $N_s=1500$, $N_z=4$, $b=0.325$ and $50$ NRG iterations 
    were performed. 
  }
  \label{fig:gfU4-III}
\end{figure}

Therfore, the extraction of the charge scale within the GWA at 
intermediate $U$ is somewhat ambiguous.
A comparison of the zero-temperature equilibrium spectral function 
of the GWA and NRG
for $U=4\Gamma$ is depicted in Fig.~\ref{fig:gfU4-III}. The low-energy feature of the GWA 
spectral function is too narrow and exhibits a rather spiky line-shape
which suggest at too low charge scale.
This is supported by the evolution of $\rho^r(0)$ as a function of 
temperature which is  shown in the 
inset of Fig.~\ref{fig:gfU4-III}. The logarithmic increase
of $\rho^r(0)$ which occurs at temperatures on the order of the relevant charge 
scale also reveals 
that the charge scale is predicted as too low in GWA compared to the NRG.
However, a considerable broadening occurs away from the Fermi level which
leads to the same FWHM for the GWA as in the NRG and 
consequently the larger $T_K^{charge}$ emerges in thermodynamic quantities like $G(T)$.

The magnetic scale $T_K^{mag}$ extracted from the GWA exhibits some
peculiar $U$-dependence. 
For small $U\apprle\Gamma$ the scale agrees with the NRG.
However, it develops a minimum at $U\approx 2.5\Gamma$ 
and then increases again for increasing $U$! This clearly indicates a failure of the 
GWA to describe magnetic properties for intermediate and large interactions.
Since the GWA effective moments $\mu_{eff}^2$ 
show universality as functions of the dimensionless temperature $t=T/T^{mag}_K$ 
for low temperatures (not shown), the increase in $T_K^{mag}$ implies
a too strong screening of magnetic moments.  
The effective Coulomb interaction is over-screened by 
$W$ (see Eq.~(\ref{eq:WrGW})).  
The electrons remain itinerant even at rather large $U$, and the GWA fails to
capture the atomic limit. Therefore, the magnetic screening
scale $T_K^{mag}$ remains large in the GWA and actually increases with $U$.

The scales extracted from the other two 
diagrammatic approximations all coincide
with the GWA for small $U\apprle \Gamma$.
For larger $U$  the $2^{nd}U$ approximation produces the same difference
as the GWA between the charge and the magnetic scale, whereas 
within the 2BA 
both scales decrease with increasing $U$, but in a polynomial rather 
than exponential fashion. 

We have established that the diagrammatic approach produces reliable results for 
interactions up to the order of the hybridization strength $U\apprle \Gamma$ 
which we will use  in the following section to  benchmark the SNRG in that regime.

\subsection{Weak correlation regime: $U \apprle \Gamma$}
\label{sec:weak-coupling}

We study the nonequilibrium properties of a symmetric and an 
asymmetric junction 
in the weakly correlated regime  $U/\Gamma\apprle 1$.
We use a very low temperature $T=0.006\Gamma$, which is
sufficiently small compared to all other scales in the problem
so it can be considered as $T=0$ with impunity.

\begin{figure}[tbp]
  \includegraphics[scale=0.65]{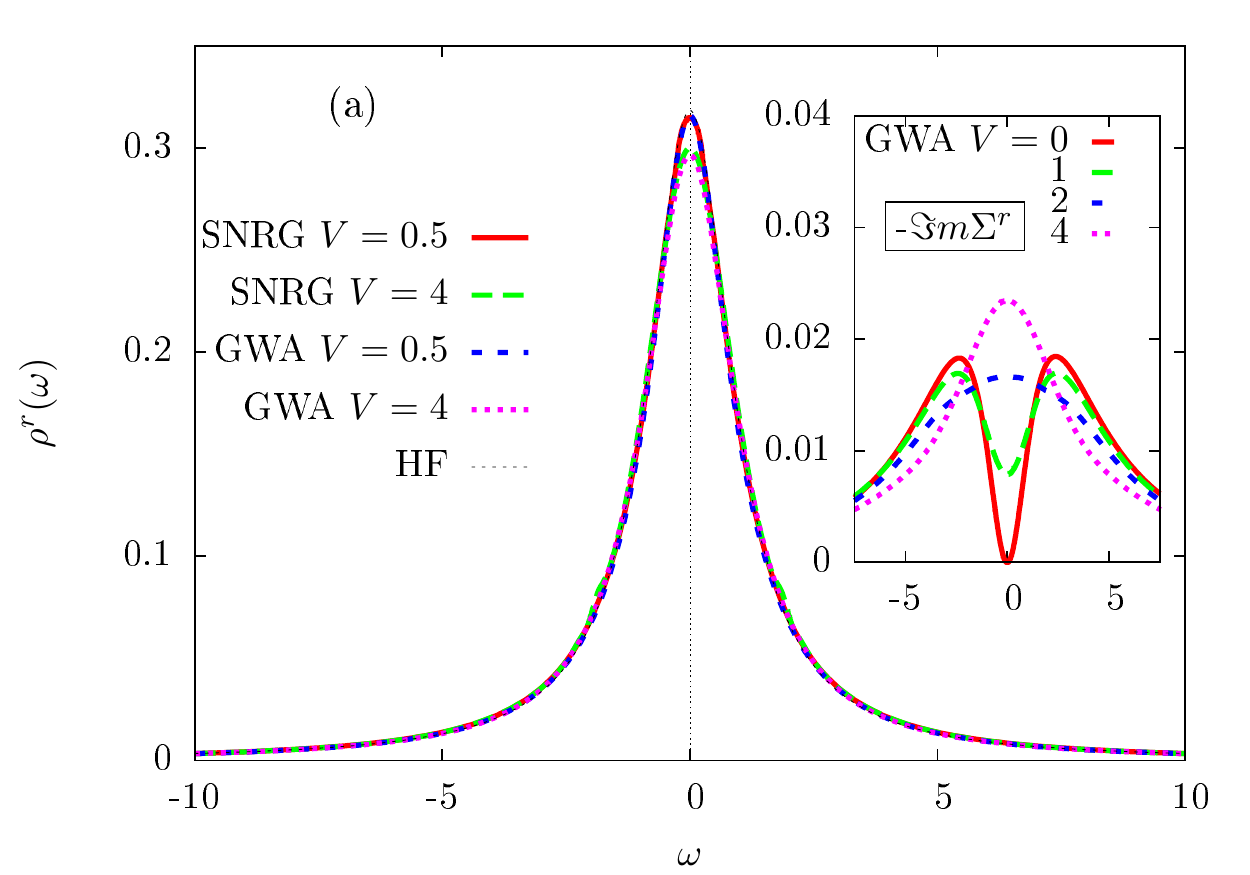}  \includegraphics[scale=0.65]{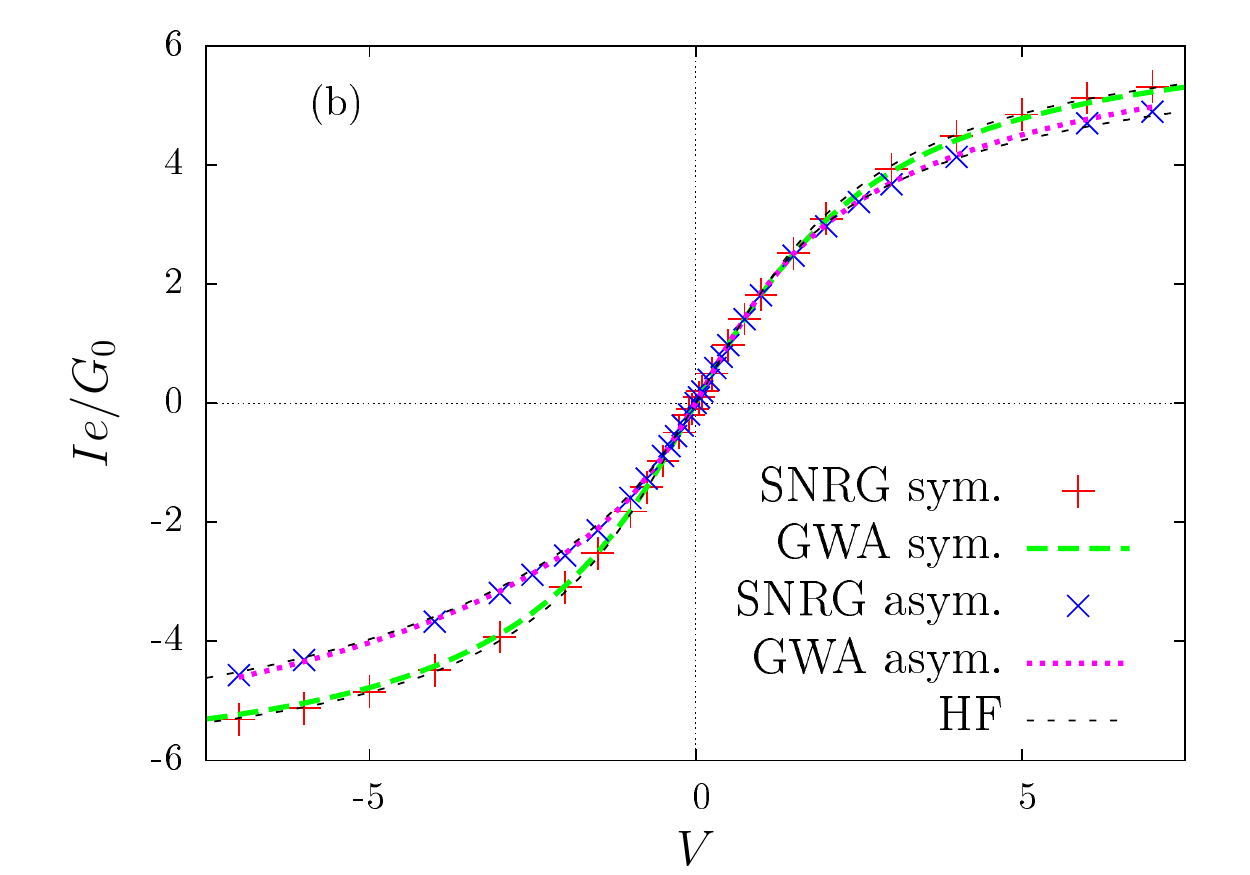}
  \caption{ (a) Spectral function of the retarded Green's function and negative imaginary part of 
    the retarded self-energy (inset) 
    of a symmetric junction with $U=-2E_d=1$ and  $\Gamma_L=\Gamma_R=0.5$ for different voltages.
    Results for the spectral functions are shown 
    for HF,  GWA and SNRG, while the self-energy is shown for the GWA only. 
    (b) Current as a function of voltage for a symmetric ($U=-2E_d=1$ and $\Gamma_L=\Gamma_R=0.5$) 
    and an asymmetric junction ( $U=1$, $E_d=-0.25$ and $\Gamma_L=4\Gamma_R=0.8$). 
    $I$ is normalized to $G_0/e=h/e$ (symmetric) and  $G_0/e=0.64h/e$ (asymmetric) and measured 
    in units of $\Gamma=1$.
    A small temperature of $T=0.006$ was used for all calculations.
    Parameters for the SNRG calculations are 
    $\Lambda=4$, $N_{s}=2200$, $N_z=2$, $b=0.65$ and 12 NRG-iterations 
    were performed.
  }
  \label{fig:weakCoupAll}
\end{figure}

For such small interactions the diagrammatic approaches and the SNRG yield identical results
for all voltages. Figure~\ref{fig:weakCoupAll}(a) shows 
the  nonequilibrium spectral function  of a symmetric junction 
in the quantum-point contact regime,
i.e.\ $U=-2E_d=\Gamma=1$ and $\Gamma_L=\Gamma_R=\Gamma/2=0.5$.
For small voltages $V/\Gamma\apprle 0.5$  the spectra 
are even indistinguishable from the Hartree-Fock (HF) result. Only 
at larger $V$ small deviations around the Fermi level 
as can be observed.

The imaginary part of the retarded self-energy $-\Im m[\Sigma^r(\w)]$ for that junction 
obtained with the 
GWA is shown in the inset of Fig.~\ref{fig:weakCoupAll}(a) for various voltages.
The overall scale of $-\Im m[\Sigma^r(\w)]$ 
is much smaller than $\Gamma$ and, therefore, the total self-energy $\Sigma^{tot}=\Sigma^r+\Delta$
is dominated by the charge-fluctuation scale $\Gamma$.
But the general influence of a finite bias voltage can already be observed here:
The quasiparticle scattering amplitude is increased by the interplay  
between the voltage-induced fluctuations
and the interaction.
The characteristic Fermi-liquid quadratic minimum in 
$-\Im m[\Sigma^r(\w)]$ at the Fermi level is destroyed with increasing 
voltage and the local Fermi liquid prevailing in equilibrium ($V=0$) 
is suppressed at large enough bias.
The evolution of the minimum in $-\Im m[\Sigma^r(\w)]$ with voltage bears some
resemblance with a temperature evolution.  The quasi-particle coherence is destroyed
by a finite voltage in a similar fashion as with increasing temperature.

The resulting IV characteristics of a symmetric and an asymmetric junction
are shown in Fig.~\ref{fig:weakCoupAll}(b) for $U=\Gamma$.
The current is normalized to $G_0/e$ and measured in units of $\Gamma$. 
The rescaled current always  saturates at $2\pi\Gamma$  
for large voltages independent of $U$ (not shown), as required  by Eq.~(\ref{eq:currentSym}).
The initial slope at zero voltage of the IV curve remains unaltered for all 
values of $U$, in accordance with the Fermi liquid 
nature of the model at small bias and zero temperature.

While a symmetric junction with symmetric coupling to the leads 
always results in symmetric spectral functions, $\rho^r(\w,V)=\rho^r(\w,-V)$, 
and antisymmetric  IV characteristics, $I(-V)=-I(V)$ (see  Eq.~(\ref{eq:currentSym})),
an asymmetric junction in combination with asymmetric coupling yields
a non-antisymmetric IV characteristics with $I(-V) \neq-I(V)$.
This is clearly visible in Fig.~\ref{fig:weakCoupAll}(b). 
The bias window  ranges from $\mu_L=-r_R^2 V$ to $\mu_R=r_L^2 V$ and is not symmetric  
around the Fermi level.
In combination with the shift of spectral weight to higher energies in $\rho^r(\w)$
--- the center of the spectral function is at $2E_d+U> 0$ ---
this leads to a smaller contribution to the current
for negative voltages.

In contrast to the spectral functions, the IV characteristics of the SNRG and GWA 
agree perfectly with the HF results for all voltages. The current 
is rather insensitive to the detailed distribution of spectral weight and
measures only the total amount  in the
bias window $[\mu_L, \mu_R]$.

The SNRG produces the correct results for small
values of the interaction $U\apprle\Gamma$, and has thus no principal 
limitations. Therefore, the expectation that it is
reliable at arbitrary  interaction strengths as well is warranted. 

\subsection{Intermediate correlation regime: $ \Gamma\apprle U \apprle 10\Gamma $}
\label{sec:inter-coupling}

As demonstrated in the previous section the interaction
plays a minor role for small $U/\Gamma$. On the other hand, 
with an odd number of electrons on the quantum dot
and  at very large $U/\Gamma$ and 
$\Gamma-U\ll E_d\ll-\Gamma$, the system develops a Kondo 
effect as $T\to 0$ (see for example Ref.\ \onlinecite{hewson:KondoProblem93} and \onlinecite{kondo40yearsSeries05}).
The SNRG was shown \cite{AndersSSnrg2008} to 
correctly describe the strongly  correlated  Kondo regime out of equilibrium. 
The enhancement of the conductance in the Coulomb-blockade region was reproduced
for small bias, and the destruction of the many-body resonance at the Fermi level with
increasing voltage has been studied.  

In this section we will focus on the intermediate interaction regime, where
correlations become increasingly important. 
Since the diagrammatic Keldysh approximations already show deficiencies  
in equilibrium --- see, for example, in Sec.~\ref{sec:equiScale} 
or Ref. \onlinecite{wangGWSIAM07} ---
discrepancies will extend to finite voltages.

\subsubsection{Nonequilibrium spectral functions of a symmetric junction}
\label{sec:spectral-inter-coupling}

\begin{figure}[tbp!]
  \includegraphics[scale=0.65]{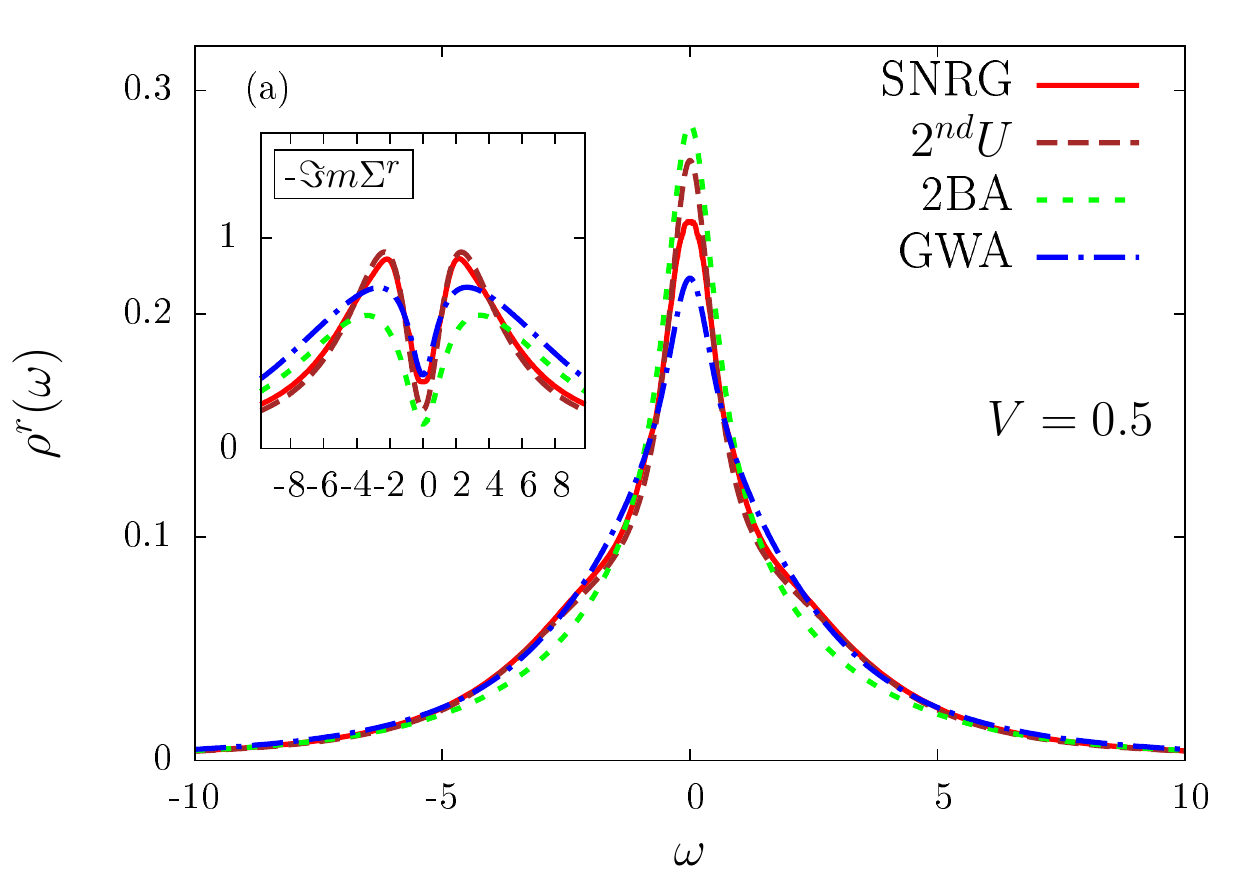}
  \\
  \includegraphics[scale=0.65]{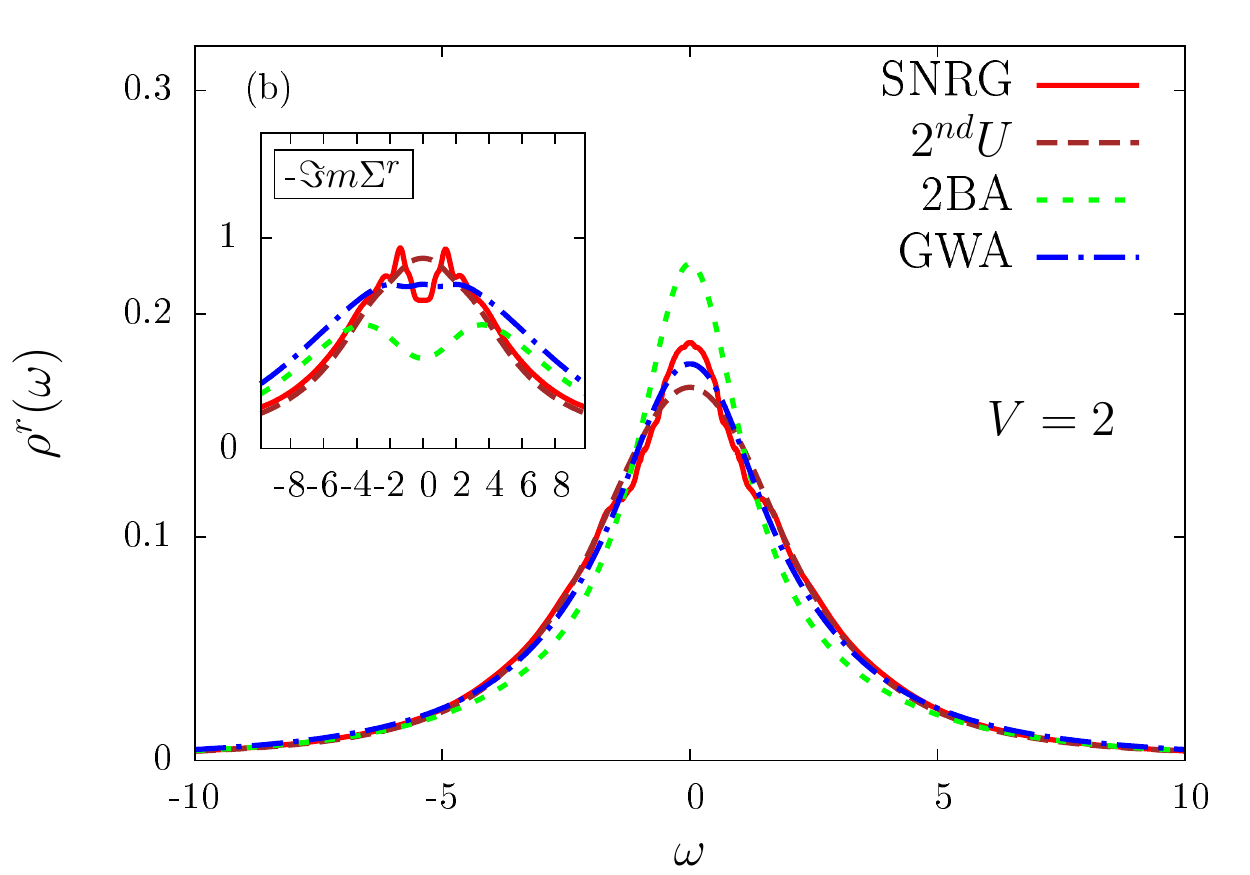}
  \\
  \includegraphics[scale=0.65]{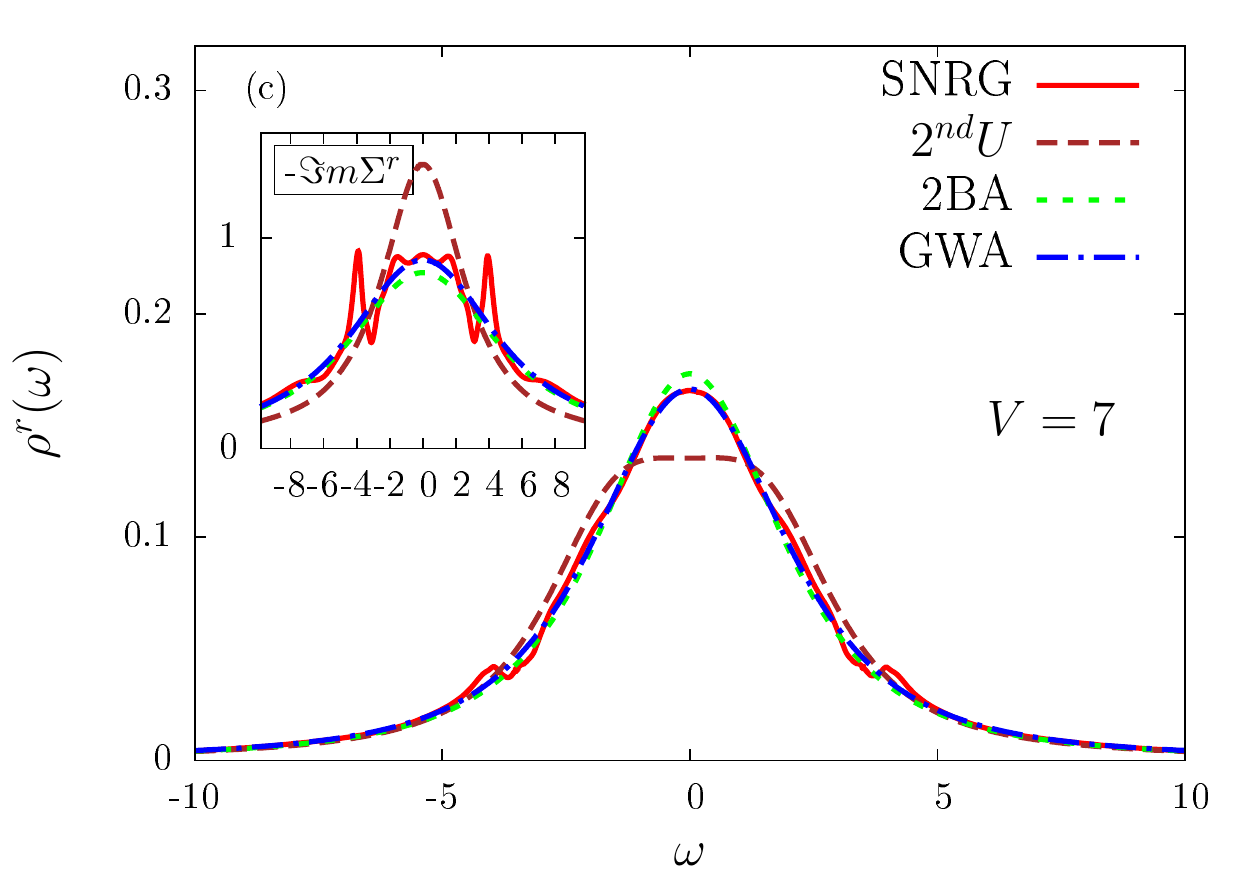}  
  \caption{Spectral functions of the SNRG and the Keldysh approaches 
    for a quantum-point contact with $\Gamma_L=\Gamma_R=0.5$ 
    and $U=-2E_d=4$ at $T=0.1$. The bias voltages are $V=0.5$ (a), $V=2$ (b) and $V=7$ (c).
    SNRG-parameters are $\Lambda=4$, $N_s=2200$, $N_z=4$, $b=0.325$ and 8 NRG-iterations were performed.
  }
  \label{fig:U4-I}
\end{figure}

The nonequilibrium spectral function for various voltages and intermediate interaction
$U=4$ is shown in Fig.~\ref{fig:U4-I}(a)-(c) for a symmetric junction ($E_d=-2$) 
with symmetric coupling to the leads ($\Gamma_L=\Gamma_R=\Gamma/2=0.5$) at $T=0.1$.
Now all diagrammatic approximations yield different results. 

At low voltages,  the SNRG and $2^{nd}U$ approximation
reproduce the slight humps at energies $\w\approx\pm U/2$ which are the first indicators
of upper and lower Hubbard satellites forming at large $U$. The GWA only produces
the broad high-energy tails, without the indication of forming separate peaks
and the 2BA completely fails to produce the enlarged spectral weight at high 
energies. 

As the voltage is raised,
the Coulomb interaction causes additional dephasing, leading to 
increasingly broadened spectra.
However, the $2^{nd}U$ approximation produces systematically too broad high-energy tails 
and an unphysical plateau around $\w=0$, which even develops a
slight dip as seen in Fig.~\ref{fig:U4-I}(c).
We attribute this to a tendency
to overestimate  the Coulomb repulsion. This might already 
be guessed from the equilibrium spectral functions, where the $2^{nd}U$
approximation unexpectedly produces the high-energy Hubbard satellites for arbitrary large Coulomb 
repulsion. These are connected to the ionic many-body states of the
isolated atom which are not expected to be described by a second-order 
perturbation theory. 
However, the analytic structure of the 
retarded self-energy,  Eq.\ (\ref{eq:SigR}), (\ref{eq:WrBA}), (\ref{eq:Pr}) and (\ref{eq:GFU0}),
has two direct consequences: 
(i) For small coupling to the leads or large $U$ it favors a 
$\Sigma^r(\w)\propto U^2/(\w+i\delta)$ behavior.
This results in a two-peak structure
with the peak-positions and widths roughly given by $\pm U$ and $\Gamma$, 
respectively. 
(Incorporating a screened and dynamic Coulomb interaction, as it is done
in the 2BA and GWA, leads to a prefactor smaller than $U^2$ 
and additional imaginary parts enter in the frequency dependence of $\Sigma^r(\w)$. The Hubbard 
satellites are then moved to lower energies and broadened.)
(ii) The Fermi functions entering Eq.~(\ref{eq:Pr}) through $G^<$ 
lead to a narrowing of the integration interval for decreasing temperature.
At zero temperature this always produces a vanishing imaginary part of the self-energy 
at the Fermi level,  
$\Im m[\Sigma^r(\w=0,T=0)]=0$,\cite{luttinger61}
given that the noninteracting propagators are non-singular at $\w=0$.
(This reasoning also holds for the 2BA and GWA).
 
The combination of (i) and (ii) gives rise to the two-peak
structure in the spectral function for large $U$ and the 
emergence of an additional peak at the Fermi level
for low temperatures
which is usually interpreted as the Kondo resonance.
But in principle there is no justification why the $2^{nd}U$ approximation
should be reliable for large values of $U$ under arbitrary conditions. 
Already for the asymmetric model in  equilibrium the phase-space argument (ii) 
does not guarantee the correct description of the low-temperature 
Fermi liquid anymore, and it is well-known that the $2^{nd}U$ approximation produces 
unphysical results.\cite{ferrerMPT87,kajuterMPT96}
Therefore, the large differences to all other 
methods at finite voltages, as it is observed here for $U/\Gamma=4$,
is not surprising.

The SNRG tends to produce additional features in the spectral function 
for large bias voltages 
at the positions of the chemical potentials of the leads, $\w\approx\mu_\alpha$.
These are the  humps visible 
for larger voltages in the curves of Fig.~\ref{fig:U4-I}.
They are artifacts of the NRG discretization and dependent on the broadening 
procedure of the NRG spectral functions.\cite{BullaCostiPruschke2007} In equilibrium, the NRG only resolves spectral information
above a cutoff frequency $|\w|>\w_c(T)$ where $\w_c$ is on the order of the temperature $T$.  
The NRG broadening parameters \cite{CostiHewsonZlatic94,BullaCostiVollhardt01,PetersPruschkeAnders2006,BullaCostiPruschke2007} 
are usually adjusted such that artefacts 
are minimized. 
Additionally the spectral function is interpolated between $-\w_c<\w<\w_c$. 
This translates itself to the present implementation of the SNRG
which does not provide spectral information in the intervals 
$I^\pm= [\mu^\pm-\w_c:\mu^\pm+\w_c]$ centered around the two chemical 
potentials. Here, $\mu^+=\mbox{max}\{\mu_L,\mu_R\}$  and $\mu^-=\mbox{min}\{\mu_L,\mu_R\}$. 
Furthermore, the time-dependent NRG introduces additional discretization errors\cite{AndersSchiller2006} 
which increase with increasing value of $U$. $z$-averaging over different 
discretizations\cite{YoshidaWithakerOliveira1990}  improves
the spectral functions and these artifacts could be removed by adjusting the 
broadening parameter depending on the voltage. In this paper, however, we keep 
the broadening\cite{CostiHewsonZlatic94,BullaCostiVollhardt01,PetersPruschkeAnders2006,BullaCostiPruschke2007} 
parameter fixed at $b=1.3/N_z$ independent of the bias and performed $z$-averaging with $N_z=2$ and $4$.

Let us focus on the different behavior of the spectral functions around $\w=0$ 
depicted in Fig.~\ref{fig:U4-I}(a)-(c).
The height $\rho^r(0)$ is reduced for increasing $V$,
and the spectral functions of the SNRG, 2BA and GWA approach each 
other and eventually coincide. 
For $V=7$, the 2BA curve is still a little higher compared  to SNRG and GWA,
but at even larger voltages (not shown) it also falls on top of the SNRG and GWA.
In contrast, the $2^{nd}U$ spectral function does not approach this large voltage limit,  
and the zero frequency value of the spectral function
is considerably reduced compared to the other approaches. This is in accord 
with an enhanced scattering amplitude at the Fermi level,  visible in the 
imaginary part of the self-energy depicted in the insets.

Upon increasing the voltage, the $2^{nd}U$ approximation does not follow a systematic 
trend since  $\rho^r(0)$
is larger  than the SNRG at low voltages and smaller at high $V$. The other approximations show 
systematic deviations as the 2BA is always larger than the SNRG, while the GWA is always
smaller.

In the present calculations, the temperature $T=0.1$ is only about on 
fifth of the equilibrium Kondo temperature for these parameter values, 
i.e.\ $T/T_K\approx 0.2$.
As discussed in Sec.~\ref{sec:equiScale}, 
the GWA produces a too small charge scale, which results in an even higher effective
temperature. 
This leads to a reduction of the spectral function at the Fermi level
in addition to the effect of the small bias. 
The  imaginary part of the self-energy 
is correspondingly too large compared to the SNRG, as can be seen in the inset of Fig.~\ref{fig:U4-I}(a).
The 2BA, on the 
other hand, overestimates the low-energy scale, which explains the 
trends in $\rho^r(0)$ and  $-\Im m[\Sigma^r(0)]$.  

Increasing the current through the junction by applying a larger bias
enhances the charge fluctuations  on the local orbital.
As already mentioned in section \ref{sec:weak-coupling},
theses additional fluctuations introduce dephasing\cite{Kehrein2005} and destroy
the coherent quasiparticles which constitute
the low-temperature Fermi liquid. The accompanying
destruction of the  characteristic quadratic minimum in $-\Im m [\Sigma^r(\w)]$ 
around $\w\approx0$ is observed in the insets. The system is driven away from the equilibrium
Fermi-liquid fixed point, and the spectral functions at the Fermi level 
decreases. At very large voltage $|V/T_K|\gg 1$ the coherent quasiparticles are 
completely suppressed, as it can be seen from the large imaginary part of the 
self-energy around $\w\approx 0$.
The spiky features in the SNRG self-energy for $V=7$ are  due to the 
aforementioned discretization errors and have no physical meaning.

\subsubsection{IV characteristics of a symmetric junction}
\label{sec:iv-inter-coupling}

\begin{figure}[tbp]
  \includegraphics[scale=0.65]{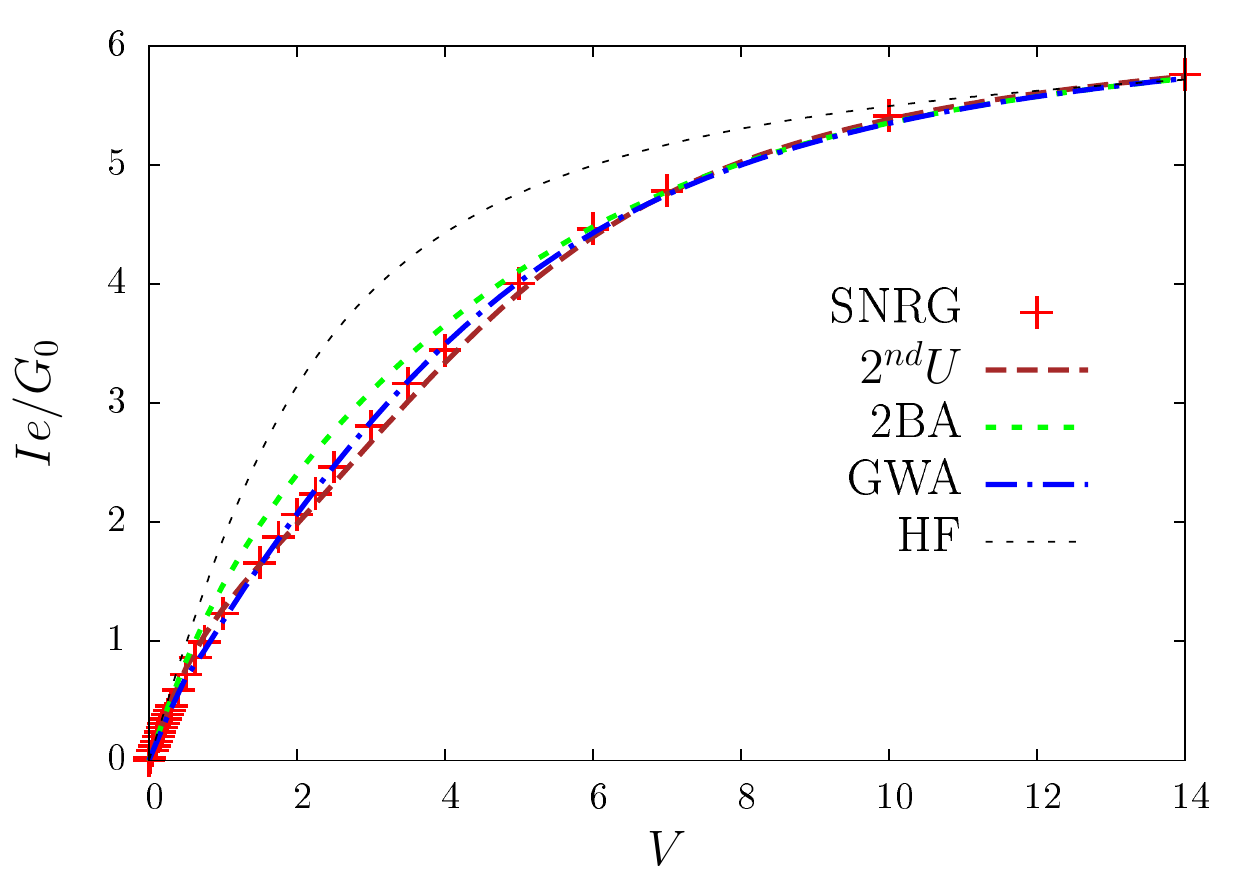}
  \caption{IV characteristics for $U=-2E_d=4$ obtained from 
    the spectral functions presented in Fig.~\ref{fig:U4-I} for a quantum-point 
    contact with $\Gamma_L=\Gamma_R=0.5$. The current is normalized to 
    $G_0/e=h/e$ and measured in units of $\Gamma=1$.
  }
  \label{fig:U4-IV} 
\end{figure}

A Coulomb interaction $U/\Gamma=4$ leads
to a reduction of the current compared to its HF value 
as  depicted in Fig.~\ref{fig:U4-IV}. 
This is characteristic for
the onset of the Coulomb blockade.  
All approaches predict this reduction but slight differences can be noticed. 
The $2^{nd}U$ approximation overestimates the Coulomb
blockade resulting in a current which is systematically smaller than the SNRG
result.
Even though the $2^{nd}U$ spectral function differs 
strongly from all other approaches for large $V$, 
this failure to describe the correct single-particle dynamics 
is concealed in the current as all approaches yield identical results.
It again shows 
the insensitivity of the current to the detailed
distribution of spectral weight in $\rho^r(\w)$.
The 2BA slightly overestimates the current for intermediate voltages, which is again
explained by the too large low-energy scale $T_K^{charge}$ and the accompanying underestimation of correlation
effects. 

The GWA current merges with the SNRG result for $V\apprge 2$, which --- together with
the satisfactory spectral function for these voltages --- indicates 
a good description of the nonequilibrium properties for intermediate to large $V$.

\subsubsection{Asymmetric junction}
\label{sec:asym-u4}

\begin{figure}[tbp]
  \includegraphics[scale=0.65]{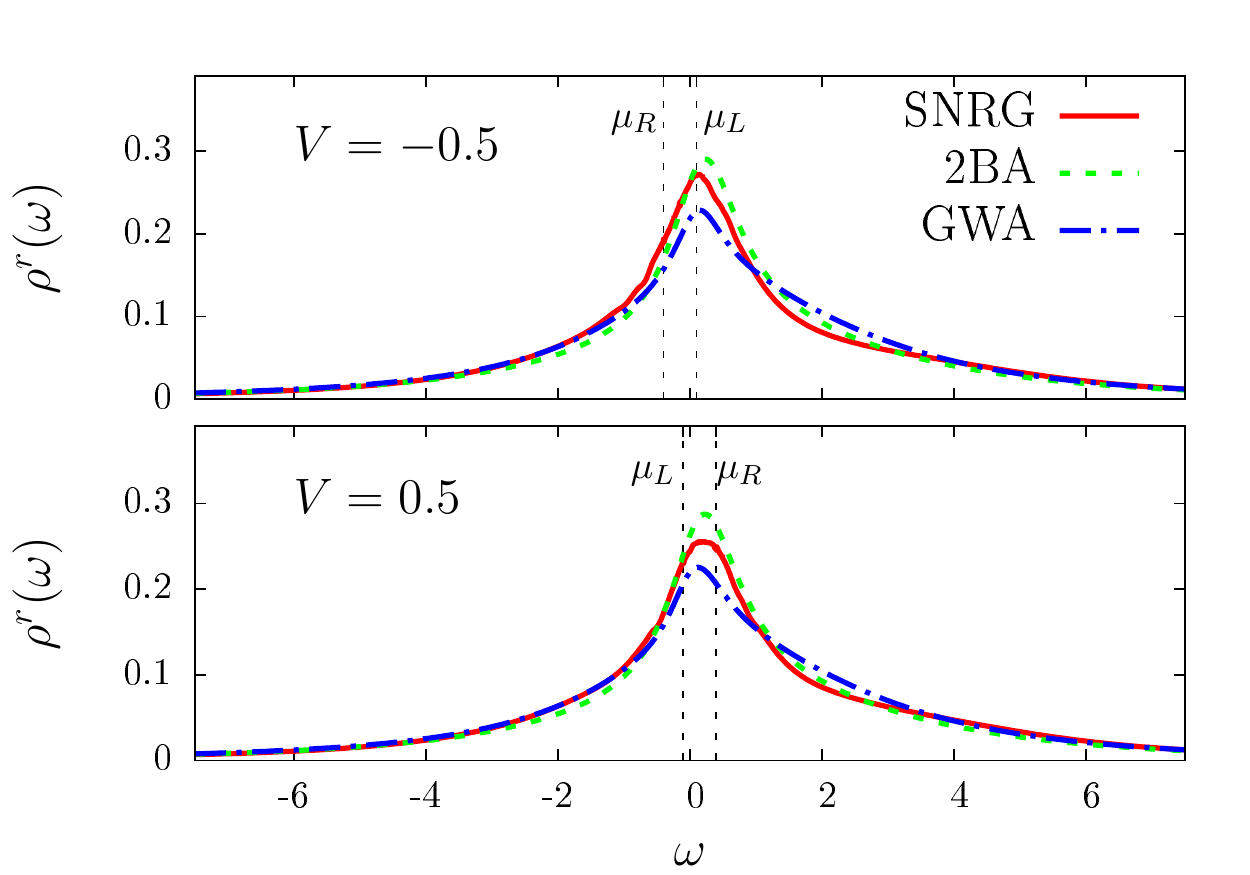}  \includegraphics[scale=0.65]{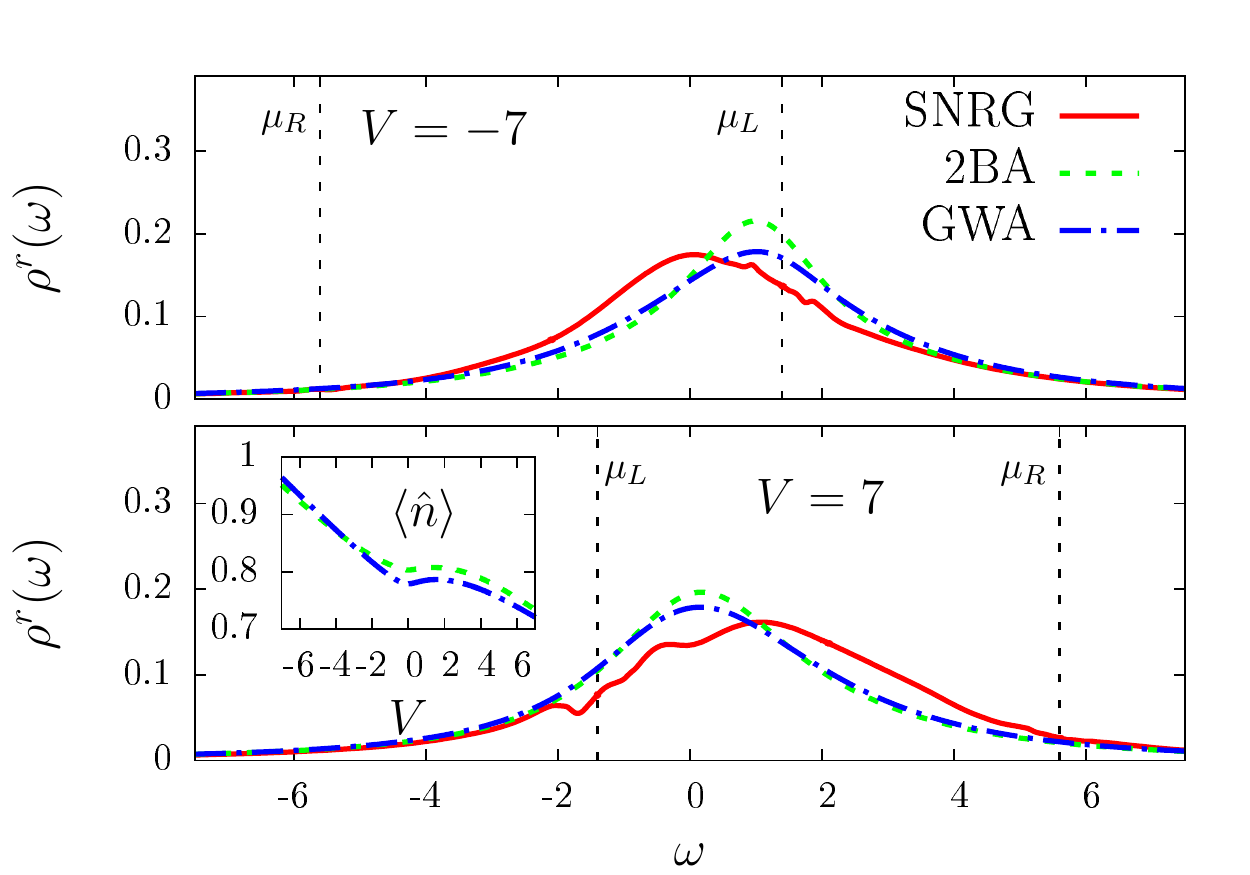}
  \caption{Spectral functions for the asymmetric junction, $U=4$, $E_d=-1$, with asymmetric coupling 
    $\Gamma_L=4\Gamma_R=0.8$ for small (upper) and large (lower) bias,  $V=\pm0.5$ and $V=\pm7$, respectively. 
    The vertical dashed lines indicate the location of the left and right chemical potentials.
    The inset shows the total occupation $\ew{\hat n}$ of the impurity as a function of the bias voltage
    obtained with the diagrammatic approaches.   
    SNRG-parameters are the same  as for Fig.~\ref{fig:U4-I}. 
  }
  \label{fig:U4AsymGF}
\end{figure}

The spectral function for 
a quantum dot with a level position $E_d=-1$, Coulomb interaction  $U=4$
and asymmetric coupling $\Gamma_L=4\Gamma_R=0.8$ is shown in Fig.~\ref{fig:U4AsymGF}.
The asymmetry between positive and
negative voltages is directly visible in the spectral functions. 

Apart from voltage-induced broadening which was already discussed in previous sections,
an additional shift of spectral weight in $\rho^r(\w)$ is observed with increasing 
$V$. 
While the SNRG moves spectral weight to higher energies for positive 
bias and towards $\w=0$ for negative $V$, 
the diagrammatic approaches produce the opposite trends.   

Due to the stronger coupling to the left lead, $\Gamma_L=4\Gamma_R$, 
the left-moving scattering states 
dominate the excitations and the spectral function of
the impurity orbital, as can be seen from Eq.~(\ref{eqn:d-LR-expand}) and~(\ref{eqn:d-alpha-cont}). 
The effective noninteracting single-particle excitation energy of an $\alpha$-mover
is given by $\Delta \e_{\alpha}= E_d-\mu_\alpha$, which 
implies almost symmetric parameters for the left-movers at the negative voltage $V=-7$ 
since then $\Delta \e_{L}=-2.4$. Therefore, the asymmetry of the
spectral function is expected to be reduced and $\rho^r(\w)$ to be closer to that of a symmetric 
junction, which is indeed observed in the SNRG. 

The diagrammatic approaches underestimate correlations in the ionic many-body states,
and the occupancy of the impurity is overestimated for negative voltages. 
It increases almost linearly with negative voltage as can be seen 
from the inset of Fig.~\ref{fig:U4AsymGF}(b).
Therefore, the Hartree shift~(\ref{eq:SigHartree}) also increases,
and spectral weight is moved towards higher energies opposite to what would
be expected from the  physical argument presented above.
As a consequence the spectral functions are strongly attracted to $\mu_L$.

The effective single-particle  excitation energy of a left-moving scattering state 
for a positive voltage $V=7$ is greater than zero, $\Delta \e_{L}=0.4$.
This produces an intermediate valence situation for the left-movers,
where correlations renormalize the effective excitation energies to even larger 
frequencies\cite{KrishWilWilson80b} and a shift of spectral weight to higher energies
results.  An additional drag of spectral weight towards the chemical potential
of the weaker coupled right lead, $\mu_R=0.8V=5.6$, is expected.  
The SNRG produces such a shift as can be seen in Fig.~\ref{fig:U4AsymGF}(b).
The diagrammatic approaches, however, underestimate the level-renormalization
in the presence of strong valence fluctuations,  a tendency  already  observable 
in equilibrium (not shown). 
Additionally, the reduced occupancy (inset) diminishes the Hartree energy which
again leads to a shift towards the stronger coupled chemical potential $\mu_L=-1.4$.

\begin{figure}[tbp]
  \includegraphics[scale=0.65]{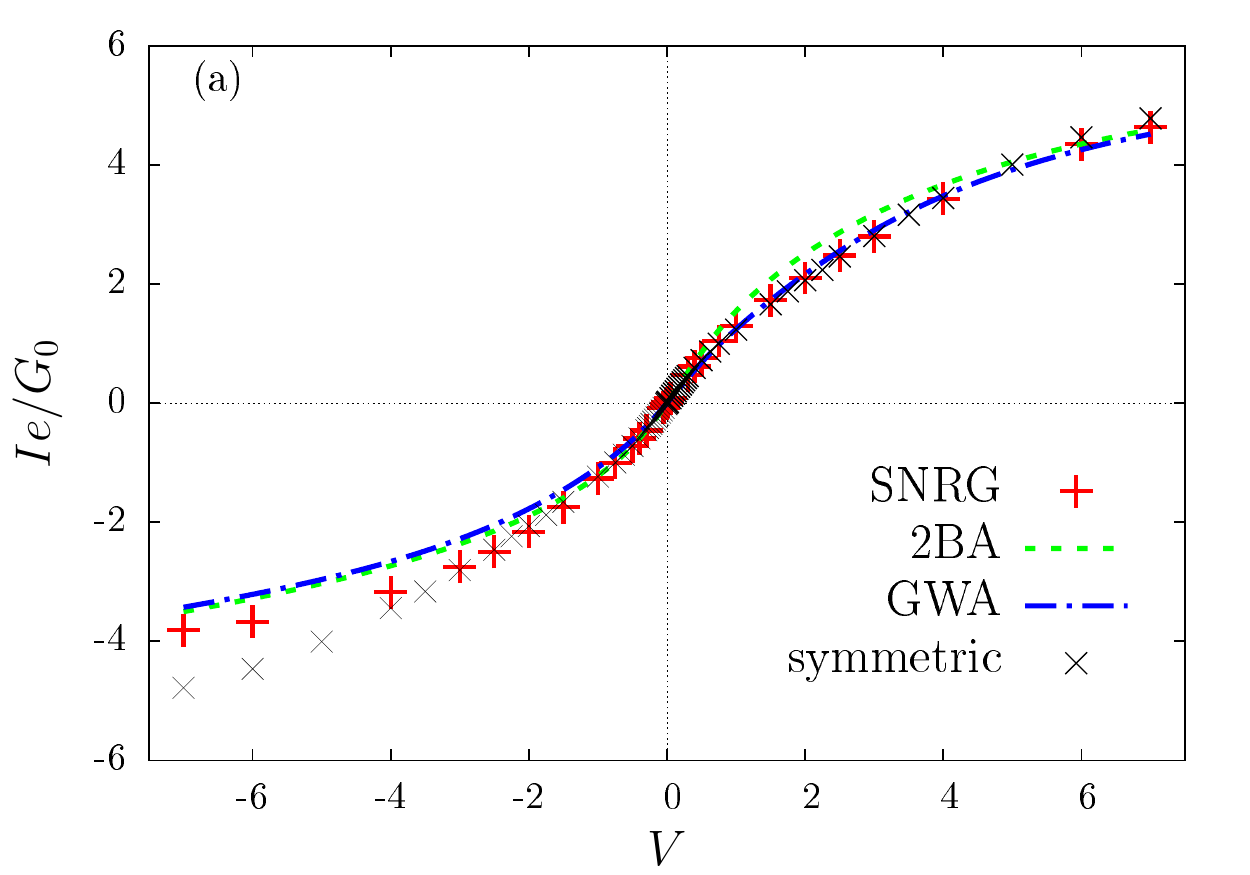} \includegraphics[scale=0.65]{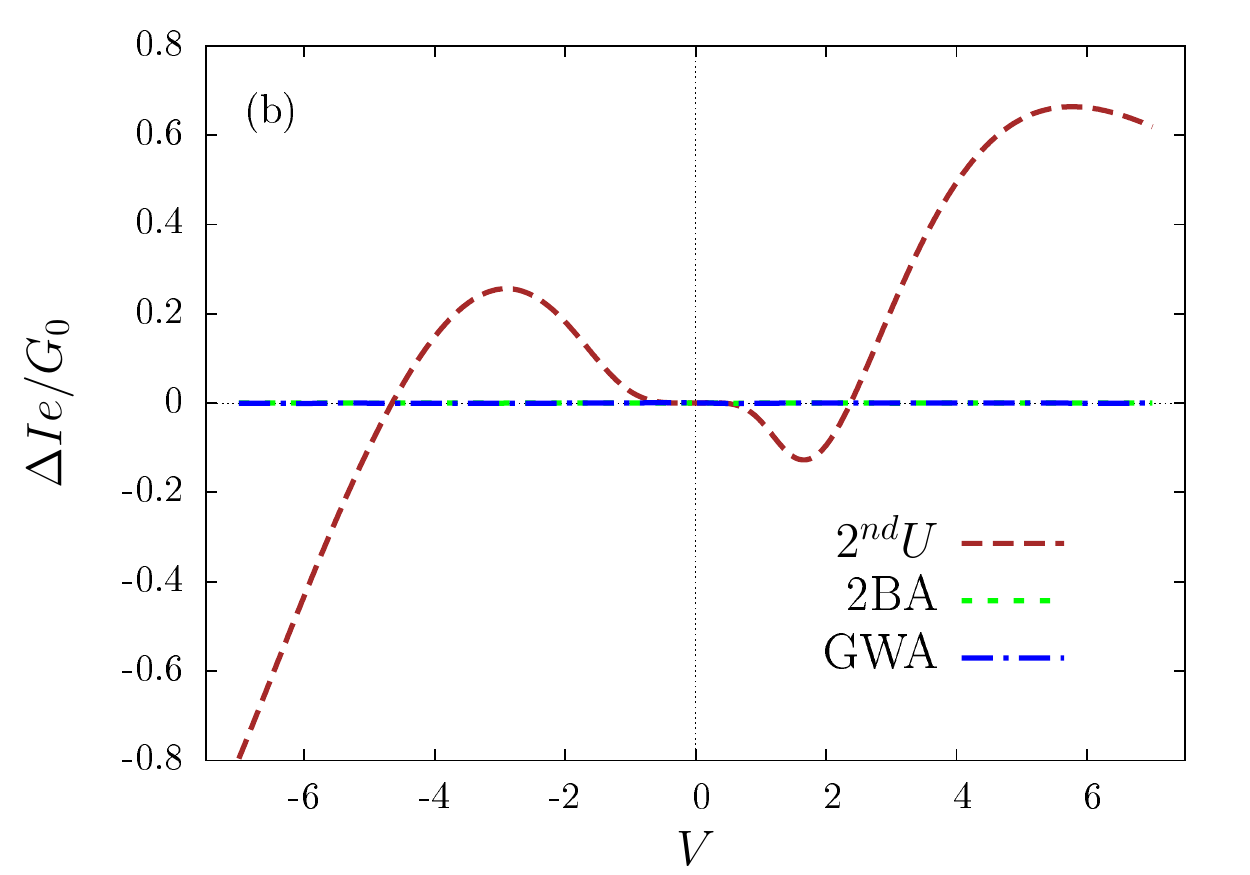}
  \caption{(a) IV characteristics of the asymmetric junction, $E_d=-1$, $U=4$
    and $\Gamma_L=4\Gamma_R=0.8$, calculated with the spectral functions depicted in 
    Fig.~\ref{fig:U4AsymGF}.  
    The result for the symmetric junction already shown 
    in Fig.~\ref{fig:U4-IV} is included for comparison.
    (b) The leakage current $\Delta I(V)$ for the Keldysh approaches
    obtained from Eq.~(\ref{eq:currentLeak}).
    The currents are normalized  to $G_0/e=0.64h/e$ and 
    measured in units of $\Gamma=1$.
  }
  \label{fig:U4AsymIV}
\end{figure}

Figure~\ref{fig:U4AsymIV} shows the IV characteristics of this
junction.
The asymmetry of $I(V)$ is clearly visible when compared to the result 
from the symmetric junction (also included in the plot). 
For $V>0$, the rescaled current is very close to its values from the 
symmetric junction and the discrepancies between the 2BA, GWA and SNRG
follow the already discussed characteristics: The 2BA underestimates correlations
and yields a slightly too large current. The GWA has the correct distribution 
of spectral weight in the bias window and produces a rather good 
estimate for the current, despite its deficiencies in the  description
of the single-particle spectra.
The failure to produce the correct shifts of spectral weights in $\rho^r(\w)$
causes the current in the diagrammatic approaches to be smaller than the SNRG
for negative voltages.

The $2^{nd}U$ approximation, however, reveals its non-conserving nature
in the violation of current conservation for this asymmetric junction.
This is illustrated in Fig.~\ref{fig:U4AsymIV}(b) displaying the leakage 
current $\Delta I$ of Eq.~(\ref{eq:currentLeak}). In contrast 
to the conserving 2BA and GWA methods and the SNRG, $\Delta I$ does not 
vanish for the $2^{nd}U$ approximation!
Thus, left and right current of Eq.~(\ref{eq:currentSym-def}) do not have the same magnitude,
i.e.\ $I_L\neq -I_R$, and the current calculated from Eq.~(\ref{eq:ss-current}) 
does not make sense, since different linear combinations $a I_L- (1-a) I_R$ ($0\leq a\leq1$)  
yield different results. Therefore, we did not include the calculated IV curves in Fig.~\ref{fig:U4AsymIV}(a).

Increasing the asymmetry further, i.e.~$\Gamma_L\gg \Gamma_R$, recovers the 
equilibrium spectral functions of a quantum dot coupled to a  single lead in all approaches
(not shown).  In the SNRG, the backscattering term 
$\hat O_{\sigma}^{back}$, Eq.~(\ref{eq:back-scattering}), is suppressed, and the model approaches
an equilibrium single-channel problem. In the diagrammatic approaches, the 
nonequilibrium conditions enter only through the effective Fermi function $f_{\mbox{eff}}$, Eq.~(\ref{eqn:f-eff}), 
which approaches its equilibrium value for $\Gamma_R\to 0$. 
In this regime, all differences in the spectral
functions of the presented approaches are given by the known discrepancies already 
present in equilibrium.

\section{Summary}
\label{sec:summary}

In the recently developed SNRG approach to open quantum systems  the scattering states of
a noninteracting quantum impurity model are used to construct 
the nonequilibrium Green's functions for the steady state at finite 
bias voltage.
We have established the reliability of the SNRG by 
benchmarking it against the diagrammatic 
Kadanoff-Baym-Keldysh approach, which becomes exact in the limit $U\to 0$.
It has been shown that the spectra and the 
current-voltage characteristics agree excellently for small Coulomb interactions
for symmetric and asymmetric junctions at arbitrary bias voltage.

For intermediate  values of $U$  we have compared the SNRG to  
three different approximations obtained from the Keldysh approach,
namely the second-order perturbation theory ($2^{nd}U$), the 
fully self-consistent second-order 
(2BA) and the GW approximation (GWA).
As correlation effects play an increasingly important
role discrepancies occur between the different methods. These were 
explained by the insufficient 
treatment of the Coulomb interaction within the diagrammatic approaches.

The Fermi liquid at zero bias voltages
is characterized by a single low-energy scale which is  
captured accurately by the SNRG, but is not properly reproduced by 
the diagrammatic approaches. 
No single low-energy scale can be extracted from the $2^{nd}U$ approximation and 
the GWA at intermediate and large $U$.
While the scale associated with charge fluctuations decreases with increasing $U$, 
the magnetic scale exhibits a qualitatively different $U$-dependency, since
it develops a minimum for intermediate $U$ and increases 
again towards larger $U$! The GWA shows a tendency to over-screen magnetic 
moments for increasing values of $U$ and fails to reproduce the atomic limit.
These deficiencies translate themselves to finite bias and explain the 
discrepancies at small to intermediate voltages.

At large bias voltages the self-consistent diagrammatic approaches (2BA and GWA)
reproduce the SNRG spectral functions for a symmetric junction, while the second-order perturbation theory 
yields an unphysical plateau around the Fermi level. 

All diagrammatic approximations and the SNRG capture 
the onset of the Coulomb blockade in the IV characteristics of the symmetric junction.
The small discrepancies are explained by the deficiencies in the treatment of
the interaction.    
However, the failure of the $2^{nd}U$ approximation to correctly describe 
the single-particle dynamics 
at large bias is masked in the current,
since there only the total
spectral weight in the bias window enters.

In contrast to the other methods, 
the $2^{nd}U$ approximation reveals its non-conserving nature by producing a 
finite leakage current for an asymmetric junction, which is unphysical.  
This raises the question about the reliability of the results obtained
within that method or 
extensions of it,\cite{yeyatiNonEqMPT93,takagiMagFieldSIAM99,fujiiPerturbKondoNeq03,fujiiNeqSIAM05,aligiaNeqSIAMPert06}
even for a symmetric junction.
They are only well-justified for cases
where $|\Sigma^r(\w)|\ll \Gamma$ for all frequencies. 

The voltage dependent redistribution of spectral 
weight for an asymmetric junction is not well-reproduced by the diagrammatic approaches. 
This has been attributed 
to too large Hartree shifts due to the wrong occupation number of the impurity
and the inaccurate renormalization of the single-particle level
in intermediate-valence situations. It leads to the underestimation of 
the current for large negative voltages.

The SNRG provides access to the description of nonequilibrium  steady-state properties
of nanoscale junctions for arbitrary Coulomb interaction and voltages.
It opens promising perspectives for future investigations, such as 
the influence of charge fluctuations when approaching the strongly-correlated 
regime or the effects of an applied magnetic field.

\section{Acknowledgments}
We are grateful to J.~E.~Han,  A.~Millis, A.~Schiller,  P.~Schmitteckert,  G.~Sch\"on, H.~Schoeller,
 P.~Werner, M.~Wegewijs and G.~Zarand for helpful discussions.
We acknowledge financial support from the Deutsche Forschungsgemeinschaft under AN  275/6-1
and supercomputer support by the NIC,
Forschungszentrum J\"ulich under project No.\ HHB000. 


%
\end{document}